%%%%%%%%%%%%%%%%%%%%% DYNKIN DIAGRAMS AND INTEGRABLE MODELS
%%%%%%%%%%%%%%%%%%%%% BASED ON LIE SUPERALGEBRAS
%%%%%%%%%%%%%%%%%%%%% J.M. EVANS AND J.O. MADSEN
%%%%%%%%%%%%%%%%%%%%%%%%%%%%%%%%%%%%%%%%%%%%%%%%%%%%%%%%%%%%%%%%%%%%
%%%%%%%%%%%%%%%%%%%%% START OF MACROS %%%%%%%%%%%%%%%%%%%%%%%%%%%%%%
%
%
%                                      Andrea PASQUINUCCI, 1988
%              PANDA.TEX               S.I.S.S.A., Trieste, Italy
%                                      (Revised 1991, Princeton, USA)
%
%--------------------------------------------------------------------
%
%    These are TEX macros. They work with PLAIN TEX (the basis
%    version of TEX). The only problem can be with the double-page
%    format since it depends on the type of software and laserwriter
%    you use to print, so I cannot guarantee that the double-page
%    format will work properly. Double-page MUST be printed in
%    LANDSCAPE orientation. (You shouldn't have troubles with fonts;
%    if you do, please let me know.)
%
%--------------------------------------------------------------------
%
%                     INTERACTIVE SECTION
%
%--------------------------------------------------------------------
%
\def\standardrisposta{s }\def\reducedrisposta{r }
\def\mplarisposta{mpla }\def\zerorisposta{z }
\def\doublerisposta{d }\def\cartarisposta{e }\def\amsrisposta{y }
\newcount\ingrandimento \newcount\sinnota \newcount\dimnota
\newcount\unoduecol \newdimen\collhsize \newdimen\tothsize
\newdimen\fullhsize \newcount\controllorisposta \sinnota=1
\newskip\infralinea  \global\controllorisposta=0
\immediate\write16 { ********  Welcome to PANDA macros (Plain TeX,
AP, 1991) ******** }
%\immediate\write16 { You'll have to answer a few questions in
%lowercase.}
%\message{>  Do you want it in double-page (d), reduced (r)
%or standard format (s) ? }\read-1 to\risposta
%\message{>  Do you want it in USA A4 (u) or EUROPEAN A4
%(e) paper size ? }\read-1 to\srisposta
%\message{>  Do you have AMSFonts 2.0 (math) fonts (y/n) ? }
%\read-1 to\arisposta
%
%--------------------------------------------------------------------
%
%             END INTERACTIVE SECTION - PAGE FORMATTING
%
%--------------------------------------------------------------------
%       The following parameters define defaults to the interactive
%       session.  At the moment I have set EUROPEAN and MATH FONTS
%\def\risposta{d } \def\srisposta{u } \def\arisposta{n }
\def\risposta{s}
\def\srisposta{e }
\def\arisposta{y }
\ifx\risposta\standardrisposta \ingrandimento=1200
\message {>> This will come out UNREDUCED << }
\dimnota=2 \unoduecol=1 \global\controllorisposta=1 \fi
\ifx\risposta\reducedrisposta \ingrandimento=1095 \dimnota=1
\unoduecol=1  \global\controllorisposta=1
\message {>> This will come out REDUCED << } \fi
\ifx\risposta\doublerisposta \ingrandimento=1000 \dimnota=2
\unoduecol=2   \message {>> You must print this in
LANDSCAPE orientation << } \global\controllorisposta=1 \fi
\ifx\risposta\mplarisposta \ingrandimento=1000 \dimnota=1
\message {>> Mod. Phys. Lett. A format << }
\unoduecol=1 \global\controllorisposta=1 \fi
\ifx\risposta\zerorisposta \ingrandimento=1000 \dimnota=2
\message {>> Zero Magnification format << }
\unoduecol=1 \global\controllorisposta=1 \fi
\ifnum\controllorisposta=0  \ingrandimento=1200
%\message {>>> ERROR IN INPUT, I ASSUME STANDARD
%UNREDUCED FORMAT <<< }  
\dimnota=2 \unoduecol=1 \fi
\magnification=\ingrandimento
%
%--------------------------------------------------------------------
%
%                        PARAMETERS SETTING
%
%  You can modify these parameters at your will (and resposability)
%--------------------------------------------------------------------
%
\newdimen\eucolumnsize \newdimen\eudoublehsize \newdimen\eudoublevsize
\newdimen\uscolumnsize \newdimen\usdoublehsize \newdimen\usdoublevsize
\newdimen\eusinglehsize \newdimen\eusinglevsize \newdimen\ussinglehsize
\newskip\standardbaselineskip \newdimen\ussinglevsize
\newskip\reducedbaselineskip \newskip\doublebaselineskip
\eucolumnsize=12.0truecm    % column h-size for european doublepage
                            % (12.0treucm default)
\eudoublehsize=25.5truecm   % sheet h-size for european duoblepage
                            % (25.5treucm default)
\eudoublevsize=6.7truein    % sheet v-size for european doublepage
                            % (6.5treuin default  or 17truecm?)
\uscolumnsize=4.4truein     % column h-size for american doublepage
                            % (4.4treuin default)
\usdoublehsize=9.4truein    % sheet h-size for american duoblepage
                            % (9.4treuin default)
\usdoublevsize=6.8truein    % sheet v-size for american doublepage
                            % (6.8treuin default)
\eusinglehsize=6.5truein    % sheet h-size for european singlepage
                            % (6.5truein default)
\eusinglevsize=24truecm     % sheet v-size for european singlepage
                            % (24truecm default)
\ussinglehsize=6.5truein    % sheet h-size for american singlepage
                            % (6.5truein default)
\ussinglevsize=8.9truein    % sheet v-size for american singlepage
                            % (8.9truein default)
\standardbaselineskip=16pt plus.2pt  % baselineskip for standard
                                     % format (16pt default)
\reducedbaselineskip=14pt plus.2pt   % baselineskip for reduced
                                     % format (14pt default)
\doublebaselineskip=12pt plus.2pt    % baselineskip for doublepage
                                     % format (12pt default)
%
%  \Portoffset and \Landoffset define the horizontal and vertical
%  offsets respectively for portrait and landscape modes. Example:
%  \def\Portoffset{\voffset=.4truein\hoffset=.125truein}
%
\def\Portoffset{}
\def\Landoffset{\voffset=-.2truein}
\ifx\risposta\mplarisposta \def\Portoffset{\hoffset=1.8truecm} \fi
%
%  \Landspec defines the \special command that sets the printer
%  to landscape mode without need to specify it directly in the
%  TeX to postscript translator (the command is site dependent).
%  Example: \def\Landspec{\special{ps: landscape}}
%
\def\Landspec{}
\tolerance=10000
\parskip=0pt plus2pt  \leftskip=0pt \rightskip=0pt
%
%   Do not modify anything of what follows
%                       (unless you know what you are doing!)
%----------------------------------------------------------------------
%
\ifx\risposta\standardrisposta \infralinea=\standardbaselineskip \fi
\ifx\risposta\reducedrisposta  \infralinea=\reducedbaselineskip \fi
\ifx\risposta\doublerisposta   \infralinea=\doublebaselineskip \fi
\ifx\risposta\mplarisposta     \infralinea=13pt \fi
\ifx\risposta\zerorisposta     \infralinea=12pt plus.2pt\fi
\ifnum\controllorisposta=0    \infralinea=\standardbaselineskip \fi
\ifx\risposta\doublerisposta   \Landoffset \else \Portoffset \fi
\ifx\risposta\doublerisposta \ifx\srisposta\cartarisposta
\tothsize=\eudoublehsize \collhsize=\eucolumnsize
\vsize=\eudoublevsize  \else  \tothsize=\usdoublehsize
\collhsize=\uscolumnsize \vsize=\usdoublevsize \fi \else
\ifx\srisposta\cartarisposta \tothsize=\eusinglehsize
\vsize=\eusinglevsize \else  \tothsize=\ussinglehsize
\vsize=\ussinglevsize \fi \collhsize=4.4truein \fi
\ifx\risposta\mplarisposta \tothsize=5.0truein
\vsize=7.8truein \collhsize=4.4truein \fi
%
%--------------------------------------------------------------------
%
%                            FONTS
%
%--------------------------------------------------------------------
%
\newcount\contaeuler \newcount\contacyrill \newcount\contaams
\font\ninerm=cmr9  \font\eightrm=cmr8  \font\sixrm=cmr6
\font\ninei=cmmi9  \font\eighti=cmmi8  \font\sixi=cmmi6
\font\ninesy=cmsy9  \font\eightsy=cmsy8  \font\sixsy=cmsy6
\font\ninebf=cmbx9  \font\eightbf=cmbx8  \font\sixbf=cmbx6
\font\ninett=cmtt9  \font\eighttt=cmtt8  \font\nineit=cmti9
\font\eightit=cmti8 \font\ninesl=cmsl9  \font\eightsl=cmsl8
\skewchar\ninei='177 \skewchar\eighti='177 \skewchar\sixi='177
\skewchar\ninesy='60 \skewchar\eightsy='60 \skewchar\sixsy='60
\hyphenchar\ninett=-1 \hyphenchar\eighttt=-1 \hyphenchar\tentt=-1
%
                 % math italic bold \bfmath
\font\tencmmib=cmmib10  \newfam\cmmibfam  \skewchar\tencmmib='177
                  % math bold (cal) symbols
\font\tencmbsy=cmbsy10  \newfam\cmbsyfam  \skewchar\tencmbsy='60
\def\scaps{\cmcsc}                 % small caps (uppercase)
\font\tencmcsc=cmcsc10  \newfam\cmcscfam
\ifnum\ingrandimento=1095

\font\capsone=cmcsc10 at 10.95pt 

\else

\font\capsone=cmcsc10 at 12pt 
\fi

\def\ttaarr{\bf}		% chapter titles' font
\def\ppaarr{\sl}		% section titles' font

%
     % inch-high caps (enormous)
%
%   AMS fonts (this works only if you have at least the 2.0
%              version of AMSFonts, otherwise say no)
%
\newfam\eufmfam \newfam\msamfam \newfam\msbmfam \newfam\eufbfam
\def\Loadeulerfonts{\global\contaeuler=1 \ifx\arisposta\amsrisposta
\font\teneufm=eufm10              %  \eufm   Gothic (or Euler)
\font\eighteufm=eufm8 \font\nineeufm=eufm9 \font\sixeufm=eufm6
\font\seveneufm=eufm7  \font\fiveeufm=eufm5
\font\teneufb=eufb10              %  \eufb   Bold Gothic (or Euler)
\font\eighteufb=eufb8 \font\nineeufb=eufb9 \font\sixeufb=eufb6
\font\seveneufb=eufb7  \font\fiveeufb=eufb5
\font\teneurm=eurm10              %  \eurm   Roman Gothic (or Euler)
\font\eighteurm=eurm8 \font\nineeurm=eurm9
\font\teneurb=eurb10              %  \eurb   Roman Bold Gothic
\font\eighteurb=eurb8 \font\nineeurb=eurb9
\font\teneusm=eusm10              %  \eusm   Slanted Capital Gothic
\font\eighteusm=eusm8 \font\nineeusm=eusm9
\font\teneusb=eusb10              %\eusb Slanted Capital Bold Gothic
\font\eighteusb=eusb8 \font\nineeusb=eusb9
\else \def\eufm{\tt} \def\eufb{\tt} \def\eurm{\tt} \def\eurb{\tt}
\def\eusm{\tt} \def\eusb{\tt}    \fi}

\def\loadamsmath{\global\contaams=1 \ifx\arisposta\amsrisposta
\font\tenmsam=msam10 \font\ninemsam=msam9 \font\eightmsam=msam8
\font\sevenmsam=msam7 \font\sixmsam=msam6 \font\fivemsam=msam5
\font\tenmsbm=msbm10 \font\ninemsbm=msbm9 \font\eightmsbm=msbm8
\font\sevenmsbm=msbm7 \font\sixmsbm=msbm6 \font\fivemsbm=msbm5
\else \def\msbm{\bf} \fi \def\Bbb{\msbm} \def\symbl{\msam} \tenpoint}
\def\loadcyrill{\global\contacyrill=1 \ifx\arisposta\amsrisposta
\font\tenwncyr=wncyr10 \font\ninewncyr=wncyr9 \font\eightwncyr=wncyr8
\font\tenwncyb=wncyr10 \font\ninewncyb=wncyr9 \font\eightwncyb=wncyr8
\font\tenwncyi=wncyr10 \font\ninewncyi=wncyr9 \font\eightwncyi=wncyr8
\else \def\cyrill{\sl} \def\cyrilb{\sl} \def\cyrili{\sl} \fi\tenpoint}
\ifx\arisposta\amsrisposta
\font\sevenex=cmex7               %  reduced math symbols
\font\eightex=cmex8  \font\nineex=cmex9
\font\ninecmmib=cmmib9   \font\eightcmmib=cmmib8
\font\sevencmmib=cmmib7 \font\sixcmmib=cmmib6
\font\fivecmmib=cmmib5   \skewchar\ninecmmib='177
\skewchar\eightcmmib='177  \skewchar\sevencmmib='177
\skewchar\sixcmmib='177   \skewchar\fivecmmib='177
%
%     CERN MODIFICATION        
%
%    The following five lines have been replaced because surya
%    does not have the necessary AMS 2 fonts
%
%\font\ninecmbsy=cmbsy9    \font\eightcmbsy=cmbsy8
%\font\sevencmbsy=cmbsy7  \font\sixcmbsy=cmbsy6
%\font\fivecmbsy=cmbsy5   \skewchar\ninecmbsy='60
%\skewchar\eightcmbsy='60  \skewchar\sevencmbsy='60
%\skewchar\sixcmbsy='60    \skewchar\fivecmbsy='60
\def\ninecmbsy{\tencmbsy}
\def\eightcmbsy{\tencmbsy}
\def\sevencmbsy{\tencmbsy}
\def\sixcmbsy{\tencmbsy}
\def\fivecmbsy{\tencmbsy}
\font\ninecmcsc=cmcsc9    \font\eightcmcsc=cmcsc8     \else
\def\cmmib{\fam\cmmibfam\tencmmib}\textfont\cmmibfam=\tencmmib
\scriptfont\cmmibfam=\tencmmib \scriptscriptfont\cmmibfam=\tencmmib
\def\cmbsy{\fam\cmbsyfam\tencmbsy} \textfont\cmbsyfam=\tencmbsy
\scriptfont\cmbsyfam=\tencmbsy \scriptscriptfont\cmbsyfam=\tencmbsy
\scriptfont\cmcscfam=\tencmcsc \scriptscriptfont\cmcscfam=\tencmcsc
\def\cmcsc{\fam\cmcscfam\tencmcsc} \textfont\cmcscfam=\tencmcsc \fi
\catcode`@=11
\newskip\ttglue
\gdef\tenpoint{\def\rm{\fam0\tenrm}
  \textfont0=\tenrm \scriptfont0=\sevenrm \scriptscriptfont0=\fiverm
  \textfont1=\teni \scriptfont1=\seveni \scriptscriptfont1=\fivei
  \textfont2=\tensy \scriptfont2=\sevensy \scriptscriptfont2=\fivesy
  \textfont3=\tenex \scriptfont3=\tenex \scriptscriptfont3=\tenex
  \def\mcal{\fam2 \tensy}  \def\mmit{\fam1 \teni}
  \textfont\itfam=\tenit \def\it{\fam\itfam\tenit}
  \textfont\slfam=\tensl \def\sl{\fam\slfam\tensl}
  \textfont\ttfam=\tentt \scriptfont\ttfam=\eighttt
  \scriptscriptfont\ttfam=\eighttt  \def\tt{\fam\ttfam\tentt}
  \textfont\bffam=\tenbf \scriptfont\bffam=\sevenbf
  \scriptscriptfont\bffam=\fivebf \def\bf{\fam\bffam\tenbf}
     \ifx\arisposta\amsrisposta    \ifnum\contaeuler=1
  \textfont\eufmfam=\teneufm \scriptfont\eufmfam=\seveneufm
  \scriptscriptfont\eufmfam=\fiveeufm \def\eufm{\fam\eufmfam\teneufm}
  \textfont\eufbfam=\teneufb \scriptfont\eufbfam=\seveneufb
  \scriptscriptfont\eufbfam=\fiveeufb \def\eufb{\fam\eufbfam\teneufb}
  \def\eurm{\teneurm} \def\eurb{\teneurb} \def\eusm{\teneusm}
  \def\eusb{\teneusb}    \fi    \ifnum\contaams=1
  \textfont\msamfam=\tenmsam \scriptfont\msamfam=\sevenmsam
  \scriptscriptfont\msamfam=\fivemsam \def\msam{\fam\msamfam\tenmsam}
  \textfont\msbmfam=\tenmsbm \scriptfont\msbmfam=\sevenmsbm
  \scriptscriptfont\msbmfam=\fivemsbm \def\msbm{\fam\msbmfam\tenmsbm}
     \fi      \ifnum\contacyrill=1     \def\cyrill{\tenwncyr}
  \def\cyrilb{\tenwncyb}  \def\cyrili{\tenwncyi}         \fi
  \textfont3=\tenex \scriptfont3=\sevenex \scriptscriptfont3=\sevenex
  \def\cmmib{\fam\cmmibfam\tencmmib} \scriptfont\cmmibfam=\sevencmmib
  \textfont\cmmibfam=\tencmmib  \scriptscriptfont\cmmibfam=\fivecmmib
  \def\cmbsy{\fam\cmbsyfam\tencmbsy} \scriptfont\cmbsyfam=\sevencmbsy
  \textfont\cmbsyfam=\tencmbsy  \scriptscriptfont\cmbsyfam=\fivecmbsy
  \def\cmcsc{\fam\cmcscfam\tencmcsc} \scriptfont\cmcscfam=\eightcmcsc
  \textfont\cmcscfam=\tencmcsc \scriptscriptfont\cmcscfam=\eightcmcsc
     \fi            \tt \ttglue=.5em plus.25em minus.15em
  \normalbaselineskip=12pt
  \setbox\strutbox=\hbox{\vrule height8.5pt depth3.5pt width0pt}
  \let\sc=\eightrm \let\big=\tenbig   \normalbaselines
  \baselineskip=\infralinea  \rm}
\gdef\ninepoint{\def\rm{\fam0\ninerm}
  \textfont0=\ninerm \scriptfont0=\sixrm \scriptscriptfont0=\fiverm
  \textfont1=\ninei \scriptfont1=\sixi \scriptscriptfont1=\fivei
  \textfont2=\ninesy \scriptfont2=\sixsy \scriptscriptfont2=\fivesy
  \textfont3=\tenex \scriptfont3=\tenex \scriptscriptfont3=\tenex
  \def\mcal{\fam2 \ninesy}  \def\mmit{\fam1 \ninei}
  \textfont\itfam=\nineit \def\it{\fam\itfam\nineit}
  \textfont\slfam=\ninesl \def\sl{\fam\slfam\ninesl}
  \textfont\ttfam=\ninett \scriptfont\ttfam=\eighttt
  \scriptscriptfont\ttfam=\eighttt \def\tt{\fam\ttfam\ninett}
  \textfont\bffam=\ninebf \scriptfont\bffam=\sixbf
  \scriptscriptfont\bffam=\fivebf \def\bf{\fam\bffam\ninebf}
     \ifx\arisposta\amsrisposta  \ifnum\contaeuler=1
  \textfont\eufmfam=\nineeufm \scriptfont\eufmfam=\sixeufm
  \scriptscriptfont\eufmfam=\fiveeufm \def\eufm{\fam\eufmfam\nineeufm}
  \textfont\eufbfam=\nineeufb \scriptfont\eufbfam=\sixeufb
  \scriptscriptfont\eufbfam=\fiveeufb \def\eufb{\fam\eufbfam\nineeufb}
  \def\eurm{\nineeurm} \def\eurb{\nineeurb} \def\eusm{\nineeusm}
  \def\eusb{\nineeusb}     \fi   \ifnum\contaams=1
  \textfont\msamfam=\ninemsam \scriptfont\msamfam=\sixmsam
  \scriptscriptfont\msamfam=\fivemsam \def\msam{\fam\msamfam\ninemsam}
  \textfont\msbmfam=\ninemsbm \scriptfont\msbmfam=\sixmsbm
  \scriptscriptfont\msbmfam=\fivemsbm \def\msbm{\fam\msbmfam\ninemsbm}
     \fi       \ifnum\contacyrill=1     \def\cyrill{\ninewncyr}
  \def\cyrilb{\ninewncyb}  \def\cyrili{\ninewncyi}         \fi
  \textfont3=\nineex \scriptfont3=\sevenex \scriptscriptfont3=\sevenex
  \def\cmmib{\fam\cmmibfam\ninecmmib}  \textfont\cmmibfam=\ninecmmib
  \scriptfont\cmmibfam=\sixcmmib \scriptscriptfont\cmmibfam=\fivecmmib
  \def\cmbsy{\fam\cmbsyfam\ninecmbsy}  \textfont\cmbsyfam=\ninecmbsy
  \scriptfont\cmbsyfam=\sixcmbsy \scriptscriptfont\cmbsyfam=\fivecmbsy
  \def\cmcsc{\fam\cmcscfam\ninecmcsc} \scriptfont\cmcscfam=\eightcmcsc
  \textfont\cmcscfam=\ninecmcsc \scriptscriptfont\cmcscfam=\eightcmcsc
     \fi            \tt \ttglue=.5em plus.25em minus.15em
  \normalbaselineskip=11pt
  \setbox\strutbox=\hbox{\vrule height8pt depth3pt width0pt}
  \let\sc=\sevenrm \let\big=\ninebig \normalbaselines\rm}
\gdef\eightpoint{\def\rm{\fam0\eightrm}
  \textfont0=\eightrm \scriptfont0=\sixrm \scriptscriptfont0=\fiverm
  \textfont1=\eighti \scriptfont1=\sixi \scriptscriptfont1=\fivei
  \textfont2=\eightsy \scriptfont2=\sixsy \scriptscriptfont2=\fivesy
  \textfont3=\tenex \scriptfont3=\tenex \scriptscriptfont3=\tenex
  \def\mcal{\fam2 \eightsy}  \def\mmit{\fam1 \eighti}
  \textfont\itfam=\eightit \def\it{\fam\itfam\eightit}
  \textfont\slfam=\eightsl \def\sl{\fam\slfam\eightsl}
  \textfont\ttfam=\eighttt \scriptfont\ttfam=\eighttt
  \scriptscriptfont\ttfam=\eighttt \def\tt{\fam\ttfam\eighttt}
  \textfont\bffam=\eightbf \scriptfont\bffam=\sixbf
  \scriptscriptfont\bffam=\fivebf \def\bf{\fam\bffam\eightbf}
     \ifx\arisposta\amsrisposta   \ifnum\contaeuler=1
  \textfont\eufmfam=\eighteufm \scriptfont\eufmfam=\sixeufm
  \scriptscriptfont\eufmfam=\fiveeufm \def\eufm{\fam\eufmfam\eighteufm}
  \textfont\eufbfam=\eighteufb \scriptfont\eufbfam=\sixeufb
  \scriptscriptfont\eufbfam=\fiveeufb \def\eufb{\fam\eufbfam\eighteufb}
  \def\eurm{\eighteurm} \def\eurb{\eighteurb} \def\eusm{\eighteusm}
  \def\eusb{\eighteusb}       \fi    \ifnum\contaams=1
  \textfont\msamfam=\eightmsam \scriptfont\msamfam=\sixmsam
  \scriptscriptfont\msamfam=\fivemsam \def\msam{\fam\msamfam\eightmsam}
  \textfont\msbmfam=\eightmsbm \scriptfont\msbmfam=\sixmsbm
  \scriptscriptfont\msbmfam=\fivemsbm \def\msbm{\fam\msbmfam\eightmsbm}
     \fi       \ifnum\contacyrill=1     \def\cyrill{\eightwncyr}
  \def\cyrilb{\eightwncyb}  \def\cyrili{\eightwncyi}         \fi
  \textfont3=\eightex \scriptfont3=\sevenex \scriptscriptfont3=\sevenex
  \def\cmmib{\fam\cmmibfam\eightcmmib}  \textfont\cmmibfam=\eightcmmib
  \scriptfont\cmmibfam=\sixcmmib \scriptscriptfont\cmmibfam=\fivecmmib
  \def\cmbsy{\fam\cmbsyfam\eightcmbsy}  \textfont\cmbsyfam=\eightcmbsy
  \scriptfont\cmbsyfam=\sixcmbsy \scriptscriptfont\cmbsyfam=\fivecmbsy
  \def\cmcsc{\fam\cmcscfam\eightcmcsc} \scriptfont\cmcscfam=\eightcmcsc
  \textfont\cmcscfam=\eightcmcsc \scriptscriptfont\cmcscfam=\eightcmcsc
     \fi             \tt \ttglue=.5em plus.25em minus.15em
  \normalbaselineskip=9pt
  \setbox\strutbox=\hbox{\vrule height7pt depth2pt width0pt}
  \let\sc=\sixrm \let\big=\eightbig \normalbaselines\rm }
\gdef\tenbig#1{{\hbox{$\left#1\vbox to8.5pt{}\right.\n@space$}}}
\gdef\ninebig#1{{\hbox{$\textfont0=\tenrm\textfont2=\tensy
   \left#1\vbox to7.25pt{}\right.\n@space$}}}
\gdef\eightbig#1{{\hbox{$\textfont0=\ninerm\textfont2=\ninesy
   \left#1\vbox to6.5pt{}\right.\n@space$}}}
 %for 10-pt math in 9-pt territory
\def\alternativefont#1#2{\ifx\arisposta\amsrisposta \relax \else
\xdef#1{#2} \fi}
\global\contaeuler=0 \global\contacyrill=0 \global\contaams=0
%
%--------------------------------------------------------------------
%
%                            MACROS
%
%--------------------------------------------------------------------
%
\newbox\fotlinebb \newbox\hedlinebb \newbox\leftcolumn
\gdef\makeheadline{\vbox to 0pt{\vskip-22.5pt
     \fullline{\vbox to8.5pt{}\the\headline}\vss}\nointerlineskip}
\gdef\makehedlinebb{\vbox to 0pt{\vskip-22.5pt
     \fullline{\vbox to8.5pt{}\copy\hedlinebb\hfil
     \line{\hfill\the\headline\hfill}}\vss} \nointerlineskip}
\gdef\makefootline{\baselineskip=24pt \fullline{\the\footline}}
\gdef\makefotlinebb{\baselineskip=24pt
    \fullline{\copy\fotlinebb\hfil\line{\hfill\the\footline\hfill}}}
\gdef\doubleformat{\shipout\vbox{\Landspec\makehedlinebb
     \fullline{\box\leftcolumn\hfil\columnbox}\makefotlinebb}
     \advancepageno}
\gdef\columnbox{\leftline{\pagebody}}
\gdef\line#1{\hbox to\hsize{\hskip\leftskip#1\hskip\rightskip}}
\gdef\fullline#1{\hbox to\fullhsize{\hskip\leftskip{#1}%
\hskip\rightskip}}
\gdef\footnote#1{\let\@sf=\empty
         \ifhmode\edef\#sf{\spacefactor=\the\spacefactor}\/\fi
         #1\@sf\vfootnote{#1}}
\gdef\vfootnote#1{\insert\footins\bgroup
         \ifnum\dimnota=1  \eightpoint\fi
         \ifnum\dimnota=2  \ninepoint\fi
         \ifnum\dimnota=0  \tenpoint\fi
         \interlinepenalty=\interfootnotelinepenalty
         \splittopskip=\ht\strutbox
         \splitmaxdepth=\dp\strutbox \floatingpenalty=20000
         \leftskip=\oldssposta \rightskip=\olddsposta
         \spaceskip=0pt \xspaceskip=0pt
         \ifnum\sinnota=0   \textindent{#1}\fi
         \ifnum\sinnota=1   \item{#1}\fi
         \footstrut\futurelet\next\fo@t}
\gdef\fo@t{\ifcat\bgroup\noexpand\next \let\next\f@@t
             \else\let\next\f@t\fi \next}
\gdef\f@@t{\bgroup\aftergroup\@foot\let\next}
\gdef\f@t#1{#1\@foot} \gdef\@foot{\strut\egroup}
\gdef\footstrut{\vbox to\splittopskip{}}
\skip\footins=\bigskipamount
\count\footins=1000  \dimen\footins=8in
\catcode`@=12
\tenpoint
\ifnum\unoduecol=1 \hsize=\tothsize   \fullhsize=\tothsize \fi
\ifnum\unoduecol=2 \hsize=\collhsize  \fullhsize=\tothsize \fi
\global\let\lrcol=L      \ifnum\unoduecol=1
\output{\plainoutput{\ifnum\tipbnota=2 \clearnmbnota\fi}} \fi
\ifnum\unoduecol=2 \output{\if L\lrcol
     \global\setbox\leftcolumn=\columnbox
     \global\setbox\fotlinebb=\line{\hfill\the\footline\hfill}
     \global\setbox\hedlinebb=\line{\hfill\the\headline\hfill}
     \advancepageno  \global\let\lrcol=R
     \else  \doubleformat \global\let\lrcol=L \fi
     \ifnum\outputpenalty>-20000 \else\dosupereject\fi
     \ifnum\tipbnota=2\clearnmbnota\fi }\fi
\def\ifdoublepage{\ifnum\unoduecol=2 }
\gdef\yespagenumbers{\footline={\hss\tenrm\folio\hss}}
\gdef\ciao{ \ifnum\fdefcontre=1 \endfdef\fi
     \par\vfill\supereject \ifnum\unoduecol=2
     \if R\lrcol  \headline={}\nopagenumbers\null\vfill\eject
     \fi\fi \end}

\newskip\olddsposta \newskip\oldssposta
\global\oldssposta=\leftskip \global\olddsposta=\rightskip

\def\filldots{\leaders\hbox to 1em{\hss.\hss}\hfill}
\def\inquadrb#1 {\vbox {\hrule  \hbox{\vrule \vbox {\vskip .2cm
    \hbox {\ #1\ } \vskip .2cm } \vrule  }  \hrule} }
 \def\newline{\hfil\break}
\def\jump{\vskip\baselineskip} \newskip\iinnffrr
\def\sjump{\iinnffrr=\baselineskip
          \divide\iinnffrr by 2 \vskip\iinnffrr}
\def\bjump{\vskip\baselineskip \vskip\baselineskip}
\newcount\nmbnota  \def\clearnmbnota{\global\nmbnota=0}
\newcount\tipbnota \def\letterfootnote{\global\tipbnota=1}

\def\note#1{\global\advance\nmbnota by 1 \ifnum\tipbnota=1
    \footnote{$^{\rm\nttlett}$}{#1} \else {\ifnum\tipbnota=2
    \footnote{$^{\nttsymb}$}{#1}
    \else\footnote{$^{\the\nmbnota}$}{#1}\fi}\fi}
\def\nttlett{\ifcase\nmbnota \or a\or b\or c\or d\or e\or f\or
g\or h\or i\or j\or k\or l\or m\or n\or o\or p\or q\or r\or
s\or t\or u\or v\or w\or y\or x\or z\fi}
\def\nttsymb{\ifcase\nmbnota \or\dag\or\sharp\or\ddag\or\star\or
\natural\or\flat\or\clubsuit\or\diamondsuit\or\heartsuit
\or\spadesuit\fi}   \clearnmbnota
\def\numberfootnote{\global\tipbnota=0} \numberfootnote
\def\setnote#1{\expandafter\xdef\csname#1\endcsname{
\ifnum\tipbnota=1 {\rm\nttlett} \else {\ifnum\tipbnota=2
{\nttsymb} \else \the\nmbnota\fi}\fi} }
\newcount\nbmfig  \def\clearnbmfig{\global\nbmfig=0}
\gdef\figure{\global\advance\nbmfig by 1
      {\rm fig. \the\nbmfig}}   \clearnbmfig
\def\setfig#1{\expandafter\xdef\csname#1\endcsname{fig. \the\nbmfig}}
 \def\endformula{\eqno\numero $$}
 \def\efr{\endformula}
\newcount\frmcount \def\clearfrmcount{\global\frmcount=0}
\def\numero{\global\advance\frmcount by 1   \ifnum\indappcount=0
  {\ifnum\cpcount <1 {\hbox{\rm (\the\frmcount )}}  \else
  {\hbox{\rm (\the\cpcount .\the\frmcount )}} \fi}  \else
  {\hbox{\rm (\applett .\the\frmcount )}} \fi}
\def\nameformula#1{\global\advance\frmcount by 1%
\ifnum\draftnum=0  {\ifnum\indappcount=0%
{\ifnum\cpcount<1\xdef\spzzttrra{(\the\frmcount )}%
\else\xdef\spzzttrra{(\the\cpcount .\the\frmcount )}\fi}%
\else\xdef\spzzttrra{(\applett .\the\frmcount )}\fi}%
\else\xdef\spzzttrra{(#1)}\fi%
\expandafter\xdef\csname#1\endcsname{\spzzttrra}
\eqno \hbox{\rm\spzzttrra} $$}
\def\nfr{\nameformula}    
\def\nameali#1{\global\advance\frmcount by 1%
\ifnum\draftnum=0  {\ifnum\indappcount=0%
{\ifnum\cpcount<1\xdef\spzzttrra{(\the\frmcount )}%
\else\xdef\spzzttrra{(\the\cpcount .\the\frmcount )}\fi}%
\else\xdef\spzzttrra{(\applett .\the\frmcount )}\fi}%
\else\xdef\spzzttrra{(#1)}\fi%
\expandafter\xdef\csname#1\endcsname{\spzzttrra}
  \hbox{\rm\spzzttrra} }      \clearfrmcount
\newcount\cpcount \def\clearcpcount{\global\cpcount=0}
\newcount\subcpcount \def\clearsubcpcount{\global\subcpcount=0}
\newcount\appcount \def\clearappcount{\global\appcount=0}
\newcount\indappcount \def\clearindappcount{\indappcount=0}
\newcount\sottoparcount 

\def\applett{\ifcase\appcount  \or {A}\or {B}\or {C}\or
{D}\or {E}\or {F}\or {G}\or {H}\or {I}\or {J}\or {K}\or {L}\or
{M}\or {N}\or {O}\or {P}\or {Q}\or {R}\or {S}\or {T}\or {U}\or
{V}\or {W}\or {X}\or {Y}\or {Z}\fi    \ifnum\appcount<0
\immediate\write16 {Panda ERROR - Appendix: counter "appcount"
out of range}\fi  \ifnum\appcount>26  \immediate\write16 {Panda
ERROR - Appendix: counter "appcount" out of range}\fi}
\clearappcount  \clearindappcount \newcount\connttrre
\def\clearconnttrre{\global\connttrre=0} \newcount\countref
\def\clearcountref{\global\countref=0} \clearcountref
\def\chapter#1{\global\advance\cpcount by 1 \clearfrmcount
                 \goodbreak\null\vbox{\jump\nobreak
                 \clearsubcpcount\clearindappcount
                 \itemitem{\ttaarr\the\cpcount .\qquad}{\ttaarr #1}
                 \par\nobreak\jump\sjump}\nobreak}
\def\section#1{\global\advance\subcpcount by 1 \goodbreak\null
               \vbox{\sjump\nobreak\ifnum\indappcount=0
                 {\ifnum\cpcount=0 {\itemitem{\ppaarr
               .\the\subcpcount\quad\enskip\ }{\ppaarr #1}\par} \else
                 {\itemitem{\ppaarr\the\cpcount .\the\subcpcount\quad
                  \enskip\ }{\ppaarr #1} \par}  \fi}
                \else{\itemitem{\ppaarr\applett .\the\subcpcount\quad
                 \enskip\ }{\ppaarr #1}\par}\fi\nobreak\jump}\nobreak}
\clearsubcpcount
\def\appendix#1{\global\advance\appcount by 1 \clearfrmcount
                  \goodbreak\null\vbox{\jump\nobreak
                  \global\advance\indappcount by 1 \clearsubcpcount
          \itemitem{ }{\hskip-40pt\ttaarr Appendix\ #1}
%         \itemitem{ }{\hskip-40pt\ttaarr Appendix\ \applett :\ #1}
%                  \itemitem{\ttaarr App.\applett\ }{\ttaarr #1}
             \nobreak\jump\sjump}\nobreak}
\clearappcount \clearindappcount
\def\references{\goodbreak\null\vbox{\jump\nobreak
   \itemitem{}{\ttaarr References} \nobreak\jump\sjump}\nobreak}

\clearcpcount\clearcountref

\def\setchap#1{\ifnum\indappcount=0{\ifnum\subcpcount=0%
\xdef\spzzttrra{\the\cpcount}%
\else\xdef\spzzttrra{\the\cpcount .\the\subcpcount}\fi}
\else{\ifnum\subcpcount=0 \xdef\spzzttrra{\applett}%
\else\xdef\spzzttrra{\applett .\the\subcpcount}\fi}\fi
\expandafter\xdef\csname#1\endcsname{\spzzttrra}}
\newcount\draftnum \newcount\ppora   \newcount\ppminuti
\global\ppora=\time   \global\ppminuti=\time
\global\divide\ppora by 60  \draftnum=\ppora
\multiply\draftnum by 60    \global\advance\ppminuti by -\draftnum
\def\droggi{\number\day /\number\month /\number\year\ \the\ppora
:\the\ppminuti}     \global\draftnum=0
\def\draftcomment#1{\ifnum\draftnum=0 \relax \else
{\ {\bf ***}\ #1\ {\bf ***}\ }\fi} 
%
%     Maximum number of references = 200
%     boxes 50 -> 250 reserved for references
%
\catcode`@=11
\gdef\Ref#1{\expandafter\ifx\csname @rrxx@#1\endcsname\relax%
{\global\advance\countref by 1    \ifnum\countref>200
\immediate\write16 {Panda ERROR - Ref: maximum number of references
exceeded}  \expandafter\xdef\csname @rrxx@#1\endcsname{0}\else
\expandafter\xdef\csname @rrxx@#1\endcsname{\the\countref}\fi}\fi
\ifnum\draftnum=0 \csname @rrxx@#1\endcsname \else#1\fi}
\gdef\beginref{\ifnum\draftnum=0  \gdef\Rref{\fairef}
\gdef\endref{\scriviref} \else\relax\fi
\ifx\risposta\mplarisposta \ninepoint \fi
\parskip 2pt plus.2pt \baselineskip=12pt}
\def\Reflab#1{[#1]} \gdef\Rref#1#2{\item{\Reflab{#1}}{#2}}
\gdef\endref{\relax}  \newcount\conttemp
\gdef\fairef#1#2{\expandafter\ifx\csname @rrxx@#1\endcsname\relax
{\global\conttemp=0 \immediate\write16 {Panda ERROR - Ref: reference
[#1] undefined}} \else
{\global\conttemp=\csname @rrxx@#1\endcsname } \fi
\global\advance\conttemp by 50  \global\setbox\conttemp=\hbox{#2} }
\gdef\scriviref{\clearconnttrre\conttemp=50
\loop\ifnum\connttrre<\countref \advance\conttemp by 1
\advance\connttrre by 1
\item{\Reflab{\the\connttrre}}{\unhcopy\conttemp} \repeat}
\clearcountref \clearconnttrre
\catcode`@=12
\ifx\risposta\mplarisposta \def\Reflab#1{#1.} \letterfootnote \fi

\def\slashchar#1{\setbox0=\hbox{$#1$} \dimen0=\wd0
     \setbox1=\hbox{/} \dimen1=\wd1 \ifdim\dimen0>\dimen1
      \rlap{\hbox to \dimen0{\hfil/\hfil}} #1 \else
      \rlap{\hbox to \dimen1{\hfil$#1$\hfil}} / \fi}
\ifx\oldchi\undefined \let\oldchi=\chi
  \def\cchi{{\raise 1pt\hbox{$\oldchi$}}} \let\chi=\cchi \fi
  
\def\del{\partial}   

\def\frac#1#2{{\textstyle{#1 \over #2}}}

\def\half{\ifinner {\scriptstyle {1 \over 2}}\else {1 \over 2} \fi}

\def\simge{\rlap{\raise 2pt \hbox{$>$}}{\lower 2pt \hbox{$\sim$}}}
\def\simle{\rlap{\raise 2pt \hbox{$<$}}{\lower 2pt \hbox{$\sim$}}}

\def\vbig#1#2{{\vbigd@men=#2\divide\vbigd@men by 2%
\hbox{$\left#1\vbox to \vbigd@men{}\right.\n@space$}}}

%
%--------------------------------------------------------------------
%
\newcount\fdefcontre \newcount\fdefcount \newcount\indcount
\newread\filefdef  \newread\fileftmp  \newwrite\filefdef
\newwrite\fileftmp     \def\strip#1*.A {#1}
\def\futuredef#1{\beginfdef
\expandafter\ifx\csname#1\endcsname\relax%
{\immediate\write\fileftmp {#1*.A}
\immediate\write16 {Panda Warning - fdef: macro "#1" on page
\the\pageno \space undefined}
\ifnum\draftnum=0 \expandafter\xdef\csname#1\endcsname{(?)}
\else \expandafter\xdef\csname#1\endcsname{(#1)} \fi
\global\advance\fdefcount by 1}\fi   \csname#1\endcsname}

\def\beginfdef{\ifnum\fdefcontre=0
\immediate\openin\filefdef \jobname.fdef
\immediate\openout\fileftmp \jobname.ftmp
\global\fdefcontre=1  \ifeof\filefdef \immediate\write16 {Panda
WARNING - fdef: file \jobname.fdef not found, run TeX again}
\else \immediate\read\filefdef to\spzzttrra
\global\advance\fdefcount by \spzzttrra
\indcount=0      \loop\ifnum\indcount<\fdefcount
\advance\indcount by 1   \immediate\read\filefdef to\spezttrra
\immediate\read\filefdef to\sppzttrra
\edef\spzzttrra{\expandafter\strip\spezttrra}
\immediate\write\fileftmp {\spzzttrra *.A}
\expandafter\xdef\csname\spzzttrra\endcsname{\sppzttrra}
\repeat \fi \immediate\closein\filefdef \fi}
\def\endfdef{\immediate\closeout\fileftmp   \ifnum\fdefcount>0
\immediate\openin\fileftmp \jobname.ftmp
\immediate\openout\filefdef \jobname.fdef
\immediate\write\filefdef {\the\fdefcount}   \indcount=0
\loop\ifnum\indcount<\fdefcount    \advance\indcount by 1
\immediate\read\fileftmp to\spezttrra
\edef\spzzttrra{\expandafter\strip\spezttrra}
\immediate\write\filefdef{\spzzttrra *.A}
\edef\spezttrra{\string{\csname\spzzttrra\endcsname\string}}
\iwritel\filefdef{\spezttrra}
\repeat  \immediate\closein\fileftmp \immediate\closeout\filefdef
\immediate\write16 {Panda Warning - fdef: Label(s) may have changed,
re-run TeX to get them right}\fi}
\def\iwritel#1#2{\newlinechar=-1
{\newlinechar=`\ \immediate\write#1{#2}}\newlinechar=-1}
\global\fdefcontre=0 \global\fdefcount=0 \global\indcount=0
%
%--------------------------------------------------------------------
%
\null
%
%--------------------------------------------------------------------
%
%                             THE    END   (OF PANDA MACROS)
%
%--------------------------------------------------------------------
%
\input epsf.tex
\font\ca=cmbsy10 at 12 pt
\font\br=cmr10 at 12pt
%%%%%%%%%%%%%%%%% Macros to rotate figures
%   These macros allow you to rotate or flip a \TeX\ box.  Very useful for
%   sideways tables or upsidedown answers.
%
%   To use, create a box containing the information you want to rotate.
%   (An hbox or vbox will do.)  Now call \rotr\boxnum to rotate the
%   material and create a new box with the appropriate (flipped) dimensions.
%   \rotr rotates right, \rotl rotates left, \rotu turns upside down, and
%   \rotf flips.  These boxes may contain other rotated boxes.
%
\newdimen\rotdimen
\def\vspec#1{\special{ps:#1}}%  passes #1 verbatim to the output
\def\rotstart#1{\vspec{gsave currentpoint currentpoint translate
   #1 neg exch neg exch translate}}% #1 can be any origin-fixing transformation
\def\rotfinish{\vspec{currentpoint grestore moveto}}% gets back in synch
%
%   First, the rotation right. The reference point of the rotated box
%   is the lower right corner of the original box.
%
\def\rotr#1{\rotdimen=\ht#1\advance\rotdimen by\dp#1%
   \hbox to\rotdimen{\hskip\ht#1\vbox to\wd#1{\rotstart{90 rotate}%
   \box#1\vss}\hss}\rotfinish}
%
%   Next, the rotation left. The reference point of the rotated box
%   is the upper left corner of the original box.
%
\def\rotl#1{\rotdimen=\ht#1\advance\rotdimen by\dp#1%
   \hbox to\rotdimen{\vbox to\wd#1{\vskip\wd#1\rotstart{270 rotate}%
   \box#1\vss}\hss}\rotfinish}%
%
%   Upside down is simple. The reference point of the rotated box
%   is the upper right corner of the original box. (The box's height
%   should be the current font's xheight, \fontdimen5\font,
%   if you want that xheight to be at the baseline after rotation.)
%
\def\rotu#1{\rotdimen=\ht#1\advance\rotdimen by\dp#1%
   \hbox to\wd#1{\hskip\wd#1\vbox to\rotdimen{\vskip\rotdimen
   \rotstart{-1 dup scale}\box#1\vss}\hss}\rotfinish}%
%
%   And flipped end for end is pretty ysae too. We retain the baseline.
%
\def\rotf#1{\hbox to\wd#1{\hskip\wd#1\rotstart{-1 1 scale}%
   \box#1\hss}\rotfinish}%
%%%%%%%%%%% End of rotation macros
%\draftmode{Toda Theories and Superalgebras}
\loadamsmath
\nopagenumbers{\baselineskip=12pt
\rightline{\hfill hep-th/9703065}
\rightline{DAMTP/96-113} 
\rightline{ENSLAPP-A-640/97} 
\ifdoublepage \bjump\bjump\bjump\bjump\else\vfill\fi
\centerline{\capsone DYNKIN DIAGRAMS AND INTEGRABLE MODELS} 
\centerline{\capsone BASED ON LIE SUPERALGEBRAS }
\bjump
\centerline{\scaps Jonathan M.~Evans\note{
Supported by a PPARC Advanced Fellowship}
}
\sjump
%\centerline{\sl Theoretical Physics Division, CERN} 
%\centerline{\sl CH-1211, Geneva 23, Switzerland}
%\centerline{\sl and }
\centerline{\sl DAMTP, University of Cambridge}
\centerline{\sl Silver Street, Cambridge CB3 9EW, UK} 
\centerline{\tt J.M.Evans@damtp.cam.ac.uk}
\sjump
\sjump
\centerline{\scaps Jens Ole Madsen}
\sjump
\centerline{\sl Laboratoire de Physique Th\'eorique}
\centerline{\sl ENSLAPP\note{
URA 14-36 du CNRS, associ\'ee \`a l'E.N.S. de Lyon et \`a 
l'Universit\'e de Savoie}, Groupe d'Annecy}
\centerline{\sl Chemin de Bellevue, F-74941 Annecy le Vieux, France} 
\centerline{\tt madsen@lapphp.in2p3.fr}

\vfill
\ifnum\unoduecol=2 \eject\null\vfill\fi
\centerline{\capsone ABSTRACT}
\sjump
\noindent 
An analysis is given of the structure of
a general two-dimensional Toda field theory involving
bosons and fermions which is defined in terms of a set of simple roots 
for a Lie superalgebra.
It is shown that a simple root system for a superalgebra 
has two natural bosonic
root systems associated with it which can be found 
very simply using Dynkin diagrams; the 
construction is closely related to the 
question of how to recover the signs of the entries of a Cartan
matrix for a superalgebra from its Dynkin diagram. 
The significance for Toda theories is that the bosonic root systems 
correspond to the purely bosonic sector of the 
integrable model, knowledge of which can determine the bosonic part of 
the extended conformal symmetry in the theory, or its classical mass
spectrum, as appropriate.
These results are applied to some special kinds of models and their
implications are investigated for features such as 
supersymmetry, positive kinetic energy and generalized reality
conditions for the Toda fields.
As a result, some new families of integrable
theories with positive kinetic energy are constructed, 
some containing a mixture of massless and
massive degrees of freedom, others being purely massive and
supersymmetric,
involving a number of coupled sine/sinh-Gordon theories.

\sjump
\ifnum\unoduecol=2 \vfill\fi
\eject
\yespagenumbers\pageno=1
%\doublespaced

%MACROS
\def\m{\mu}
\def\F{ {\cal F} }
\def\B{ {\cal B} }
\def\G{ {\cal G} }

\def\a{ {\rm A} }
\def\b{ {\rm B} }
\def\c{ {\rm C} }
\def\d{ {\rm D} }
\def\e{ {\rm E} }
\def\f{ {\rm F} }
\def\g{ {\rm G} }
\def\u{ {\rm U} }

%% A, B, D with Upper and Lower stars
\def\au{ {\rm A} \kern -1pt \hbox{*} }
\def\bu{ {\rm B} \hbox{*} } 
\def\du{ {\rm D} \hbox{*} } 
\def\bl{ {\rm B} \lower 4pt \hbox{*} }
\def\dl{ {\rm D} \lower 4pt \hbox{*} }

\def\psib{\bar \psi}
\def\chib{\bar \chi}
\def\etab{\bar \eta}

\def\R{ {\cal R} }
\def\gle{ \kern 1.5pt \hbox{$\vphantom{x}^>$} \kern -6.5pt
\raise 0.5pt \hbox{$\vphantom{X}_<$} \kern 1.5pt }
\def\Half{ {\textstyle {1 \over 2}} }

%1
\chapter{Introduction}

Toda theories provide a uniform way of constructing both massless and massive
integrable field theories from Lie algebras 
[\Ref{LS}\vphantom{\Ref{PM},\Ref{MOP}}--\Ref{OT}]. 
In recent years these models have received considerable attention.
It is well known that those based on
finite-dimensional Lie algebras exhibit extended conformal
symmetry (see [\Ref{MS}\vphantom{\Ref{HR}}--\Ref{BS}] for reviews) 
whilst those based on affine Kac-Moody algebras are 
massive integrable field theories for which exact S-matrices can be 
found (see eg.~[\Ref{BTS}]). 
Toda theories based on Lie superalgebras
have also been considered by a number of authors 
[\Ref{O}\vphantom{\Ref{LSS}\Ref{LM1}\Ref{LM2}
\Ref{EH1}\Ref{JE}\Ref{EH2}\Ref{NM}\Ref{WI}\Ref{SHR}\Ref{EM1}
\Ref{ABL}\Ref{Ahn}\Ref{IK}\Ref{KU}\Ref{STS}\Ref{QI}}--\Ref{PPZ}].  
These integrable models contain both bosonic and fermionic fields,
but they need not be supersymmetric in general.
They have been studied principally because 
they provide field-theoretic realizations of $\cal W$-algebras,
[\Ref{EH1}-\Ref{SHR}] but there has also 
been work concerned with S-matrices for massive models
[\Ref{ABL}--\Ref{PPZ}].
Results for theories based on superalgebras are much less
complete than in the bosonic case, however.

In this paper we establish a number of results 
concerning the structure of a Toda theory based on
a general Lie superalgebra. 
We show that the bosonic sector of any
such theory consists
of a sum of bosonic Toda models whose lagrangians can contribute with positive 
or negative signs and 
we describe a simple way of reading off the 
Dynkin diagrams corresponding to these bosonic sub-theories from 
the Dynkin diagram of the original Lie superalgebra.
By applying this idea to a conformally-invariant theory  
one can easily determine the
bosonic subalgebra of the extended conformal symmetry which is responsible
for the integrability of the model [\Ref{MS}-\Ref{BS}],
while for a massive theory 
one can find the classical mass spectrum in terms of
the known mass spectra of bosonic models [\Ref{MOP},\Ref{BTS},\Ref{FLO}].
In addition to this information on the bosonic content of the theory,
we give a criterion for the decoupling of certain fermionic
degrees of freedom.
After explaining these general results, we discuss 
various classes of examples.
 
Almost all Toda models based on finite-dimensional Lie superalgebras are 
conformally invariant, in close analogy with the bosonic situation
(the exceptions are discussed in sections 4 and 6).
For models based on an infinite-dimensional superalgebra, however,
one finds behaviour which differs markedly from the bosonic case.
While bosonic affine Toda models are massive integrable field theories,
we shall show that the bosonic sector of a theory based on
an affine superalgebra can in general be a combination of 
massive and conformal bosonic Toda theories, typically including 
a number of free massless bosons as well. 
Our methods will allow us to identify those models which are 
purely massive.

Another way in which superalgebra theories differ from their bosonic
counterparts is that their kinetic energy can have indefinite signature.
While this is not necessarily a fatal flaw, particularly in
the context of conformal field theory where the study of non-unitary
models is common, it is clearly interesting to
ask which models have conventional, positive-definite 
kinetic energy.
This point was originally raised by Olshanetsky [\Ref{O}],
but in fact the class of Toda models he considered can be generalized
to allow for non-standard or {\it twisted\/} reality conditions on
the fields [\Ref{JE},\Ref{EM1}]. 
Our techniques enable one to see easily how such 
reality conditions descend to the bosonic sector of a superalgebra
theory. By extending the analysis in this way, we find 
new families of integrable field theories with positive-definite
lagrangians. 
Most of these 
contain both massless and massive degrees of freedom, but several
examples are purely
massive and supersymmetric, consisting of sine/sinh-Gordon theories
coupled together in various ways.

The Dynkin diagram construction at the heart of this paper is
intimately related to the question of how one recovers a Cartan matrix
from a Dynkin diagram for a Lie superalgebra, 
including the rather subtle issue of
reconstructing the correct relative signs for all the
entries.
We discuss this in detail in the next section.
We then move on to Toda theories proper and to the topics indicated above.
Throughout the paper we consider only {\it classical\/}, {\it abelian\/} 
Toda theories with {\it real\/} coupling constants, but we shall draw
attention at various stages to those 
theories that would be particularly interesting to investigate further
at the quantum level. 

%2 
\chapter{Simple roots and Dynkin diagrams} 

We shall be concerned with 
Lie superalgebras $\G(\R)$ defined in terms of some finite
set of simple roots $\R$. This means we consider 
precisely the superalgebras of types A, B, C, D, E,
F, G and their affine extensions, as defined by the Cartan-Kac
classification. We first recall some elementary facts concerning 
these algebras; for more details, see
[\Ref{LA}\vphantom{\Ref{GO},\Ref{Kac0},\Ref{Kac1},\Ref{Kac2},\Ref{Kac3},
\Ref{Serg},\Ref{Leur},\Ref{LSS},\Ref{FSS}}--\Ref{Dict},\Ref{LSS}].

The simple roots may be identified 
with a set of $n$ vectors $\alpha_i$ 
living in an $r$-dimensional space $V$, with $i \in \{ 1, \ldots, n
\}$ (for affine algebras we 
project onto a horizontal subspace).
Each simple root is graded 
bosonic (even) or fermionic (odd) as specified by 
the disjoint union 
$\R = \B \cup \F$ or equivalently by values of the
labels $ i \in \varepsilon $ and $i \in \tau$ respectively,
where $\varepsilon$ and $\tau$ are complementary subsets of 
$\{1, \ldots , n\}$.
We denote the span of any set of vectors ${\cal S}$ in $V$ by
$\langle {\cal S} \rangle$.
The simple roots span the space, so 
$V = \langle \R \rangle$ and $n \geq r$
(in fact $n=r$ for finite-dimensional superalgebras and
$n=r{+}1$ for infinite-dimensional superalgebras,
with the exceptions of $\a(m,m)$ and $\a(m,m)^{(1)}$ which have
$n=r{+}1$ and $n=r{+}2$ respectively.)
Every superalgebra $\G (\R)$ contains a unique maximal
bosonic subalgebra which we denote by $\G(\R)_{\rm bos}$, but 
this depends on the entire root system 
$\R$ and not just on the bosonic simple roots $\B$.

There is a natural inner-product on $V$ 
which will be denoted by a dot and which is inherited from the 
invariant inner-product on the algebra $\G(\R)$.
The inner-product is non-degenerate on $V$, by construction (since we
take projected roots for affine algebras).
For bosonic algebras the inner-product is 
positive-definite but for Lie superalgebras it is typically of
indefinite signature, 
which gives rise to some very important differences. 
If $U$ is any subspace of $V$ and 
$U^\bot$ denotes the orthogonal subspace, then 
$U \cap U^\bot \neq \{ 0 \}$ in general.
If the restriction of the inner-product to $U$ is non-degenerate,
however, then $U \cap U^\bot = \{ 0 \} $ and $V = U \oplus U^\bot$.

The precise definition of 
$\G(\R)$ can be given in terms of sets of 
generators 
$H_i$ and $E^\pm_i$ corresponding to the simple
roots.
These must satisfy the basic relations  
$$
[H_i , E^\pm_j ] = \pm k_{ij} E^\pm_j \ ,
\qquad
[E^+_i , E^-_j ] = \delta_{ij} H_i \ .
\nfr{gens}
The Cartan matrix $k_{ij}$ has dimension $n{\times}n$, rank $r$,
and it encodes all necessary information about the root system.
In these equations, no summation is implied and the bracket is
understood to be graded. 
The definition of the algebra must be completed by further conditions on the
generators known as Serre relations 
[\Ref{LA},\Ref{Kac0},\Ref{Kac1},\Ref{Kac2},\Ref{Dict}]. 

It is clear that by re-scaling the generators $H_i$ and $E_i^\pm$ we can
change the Cartan matrix $k_{ij} \rightarrow \lambda_i k_{ij}$ (no summation) 
where $\lambda_i$ are any non-zero constants. 
Using this freedom, it is always possible to make the Cartan matrix 
symmetric; it is then just the matrix of inner-products of
the simple roots, which we write as 
$k_{ij} = a_{ij} = \alpha_i \cdot \alpha_j$.
The matrix $a_{ij}$ is unique up to a permutation of the simple roots 
and an overall constant. We can take its 
entries to be integers, except for 
$\d(2,1; \alpha)$ and its affine extensions at generic values of
the parameter $\alpha$. 
To define Dynkin diagrams, it is convenient to introduce a slightly
different Cartan matrix which we write as $k_{ij} = b_{ij}= \lambda_i
a_{ij}$, where $\lambda_i > 0$.
When the entries $a_{ij}$ are integers, we can
choose $\lambda_i$ so that $b_{ij}$ are also integers which obey: 
{\bf (i)} $| b_{ii} | = 2$ or 0; {\bf (ii)} if $b_{ii} = 0$ then
$b_{ij}$ ($j = 1, \ldots , n$) have no common factor. 
These conditions fix the normalization of each row, so that 
$b_{ij}$ is unique up to an overall sign. It is also useful to note 
the properties: 
{\bf (iii)} $b_{ij}$ and $b_{ji}$ have the same sign;
{\bf (iv)} if $b_{ij} \neq b_{ji}$ then 
either $|b_{ij}| = 1$ or $|b_{ji}| = 1$.
We can define $b_{ij}$ satisfying (i), (iii) and (iv) 
for the exceptional D algebras too,
but we must replace (ii) with some other criterion for normalizing the rows
corresponding to null roots if $b_{ij}$ is to be unique. 
We emphasize that with our conventions the diagonal entries 
of $a_{ij}$ and $b_{ij}$ can be
positive or negative. There are a number of other 
choices for the Cartan matrix adopted in
[\Ref{LSS},\Ref{Kac0}-\Ref{Dict}],
many of which rescale the rows so
that the diagonal entries are non-negative, so relinquishing 
property (iii).

For any bosonic Lie algebra, the simple root system $\R$ 
is essentially unique: any two simple root systems are related
by the action of the Weyl group and the Cartan matrices coincide up
to a permutation of rows and columns [\Ref{LA},\Ref{Kac0}]. 
For a Lie superalgebra, however, it is quite possible to find 
inequivalent simple root systems $\R$ and $\R'$ 
which lead to isomorphic 
algebras $\G(\R) = \G(\R')$ despite the fact that their
Cartan matrices are completely different. 
It is possible to reconstruct all allowed
simple root systems for a given superalgebra from any one simple root system
by applications of 
transformations which generalize the reflections contained in the 
bosonic Weyl group; for more details see [\Ref{LSS},\Ref{FSS},\Ref{Dict}]. 
We should  merely be aware that 
for a given algebra there may be a number of equally valid simple
roots systems $\R$.

We now give rules for encoding the Cartan matrix 
$b_{ij}$ in a Dynkin diagram---these follow very closely the usual
bosonic conventions.
\vskip 2pt
\noindent
$\bullet$ Draw a {\it node\/} for each simple root $\alpha_i$, 
coloured white if it is bosonic ($i \in \varepsilon$), black if it is 
fermionic and non-null ($i \in \tau$ and $\alpha_i^2 \neq 0$), grey or hatched
if it is fermionic and null ($i \in \tau$ and $\alpha_i^2 = 0$).
\vskip 2pt
\noindent
$\bullet$ Draw a {\it bond\/} between the nodes 
corresponding to $i$ and $j$ 
consisting of $max ( | b_{ij} | , | b_{ji} | )$ lines, this being the
{\it strength\/} of the bond, 
with an arrow from $i$ to $j$ if $| b_{ji} | > |b_{ij} | $.
\vskip 2pt
\noindent
We have assumed here that the entries of the Cartan
matrix are integers. We can extend the rules to
encompass those members of the continuous family 
$\d(2,1;\alpha)$ and
their affine counterparts whose Cartan matrices have non-integer entries.
We proceed as before, but instead of drawing a bond consisting of a
number of lines to represent
an integer, we draw a single line and label it with the real number
$b_{ij}$ whenever necessary; see eg.~[\Ref{FSS},\Ref{Dict}]. 

The scheme above defines unambiguously how one draws a Dynkin diagram 
to represent a given Cartan matrix. 
Although our choice for the Cartan matrix itself may differ in small
ways from some
others in the literature, the Dynkin diagrams which follow from the
rules we have given agree with those which can be found in 
standard references such as [\Ref{Kac1},\Ref{Kac2},\Ref{FSS},\Ref{Dict}].
So far everything seems like a simple modification of the bosonic
situation. 
The big difference, however, is that one cannot uniquely
reconstruct the Cartan matrix for a superalgebra 
from its Dynkin diagram without giving some additional rules.
The problematical thing is how to
recover the {\it signs\/} of the entries of the Cartan matrix.
We have chosen to focus on $a_{ij}$ and $b_{ij}$ because
they are the Cartan matrices which are 
most immediately relevant for constructing integrable lagrangians,
but also because the signs of their entries can be 
recovered in a relatively straightforward way.
To explain how, it is best to describe in more
detail what
kinds of Cartan matrices and diagrams are actually allowed by the
classification of [\Ref{Kac1}-\Ref{Leur}]. 

It will be convenient to refer to the  
infinite series of superalgebras of types A, B, C, D as {\it classical\/}
(so this term is being used here in a different sense than in 
[\Ref{Kac1},\Ref{Dict}]) 
and to distinguish them from
the {\it exceptional\/} algebras $\g_2$, $\f_4$, $\e_6$, $\e_7$, $\e_8$,
$\d(2,1;\alpha)$, $\g(3)$, $\f(4)$, all their 
affine extensions, and $\d_4^{(3)}$.
The Dynkin diagrams for the finite-dimensional, 
classical bosonic Lie algebras
of types A, B, C and D are shown below in Figure 1.
\vskip 15pt

\hbox{ \kern 100pt {\br A} \kern 20pt 
\lower 3pt \hbox{\epsfxsize=135pt \epsfbox{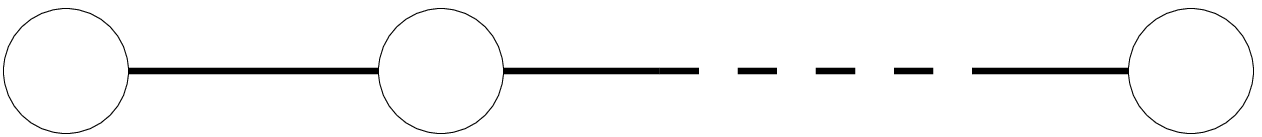}} } \vskip 25pt

\hbox{ \kern 100pt {\br B} \kern 20pt
\lower 3pt \hbox{\epsfxsize=135pt \epsfbox{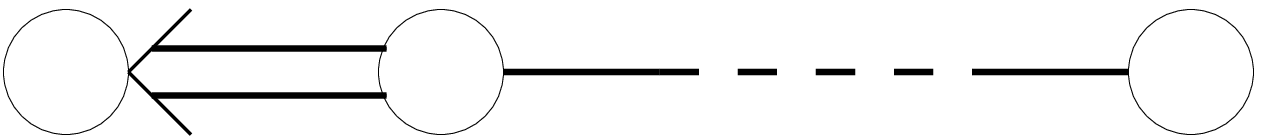}} } \vskip 25pt

\hbox{ \kern 100pt {\br C} \kern 20pt
\lower 3pt \hbox{\epsfxsize=135pt \epsfbox{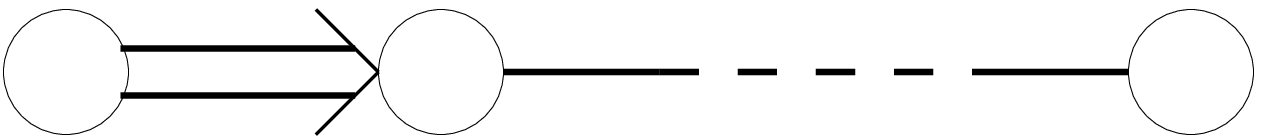}} } \vskip 20pt

\hbox{ \kern 100pt \raise 16pt \hbox{\br D} \kern 10pt
\lower 7pt \hbox{\epsfxsize=160pt \epsfbox{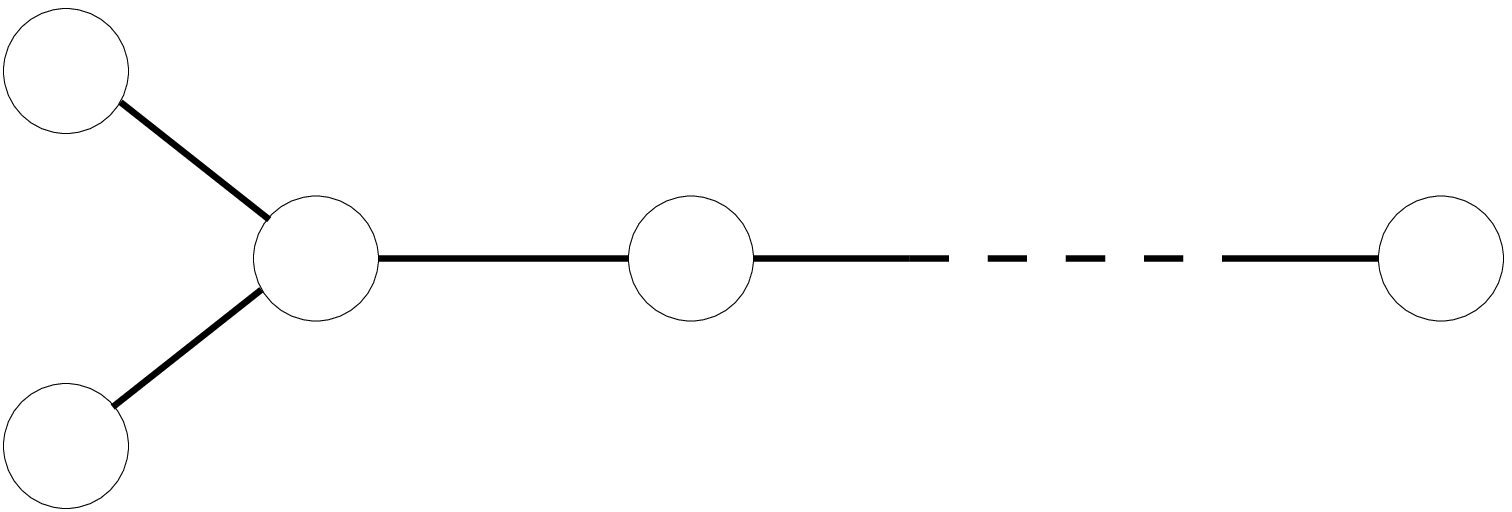}} }

\vskip 15pt
\centerline{\bf Figure 1: classical bosonic Dynkin diagrams}
\vskip 10pt
\noindent
In a similar spirit, let us introduce the symbols $\au$, $\bu$
and $\du$ corresponding to the fermionic diagrams in Figure 2.
\vskip 15pt

\hbox{ \kern 100pt {\br A*} \kern 10pt 
\lower 3pt \hbox{\epsfxsize=170pt \epsfbox{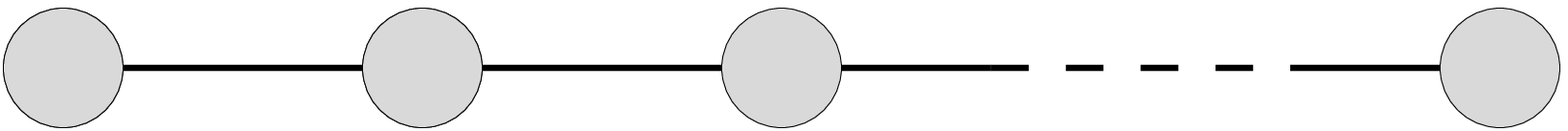}} } \vskip 25pt

\hbox{ \kern 100pt {\br B*} \kern 10pt
\lower 3pt \hbox{\epsfxsize=170pt \epsfbox{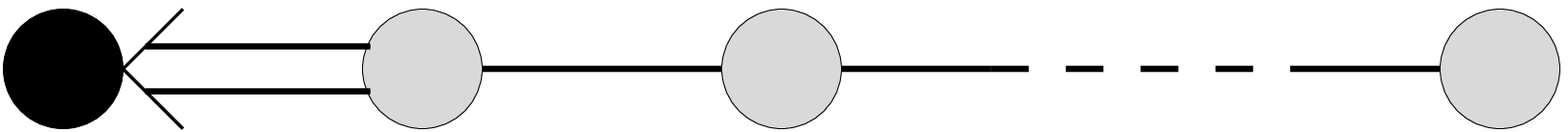}} } \vskip 20pt

\hbox{ \kern 100pt \raise 15pt \hbox{\br D*} \kern 10pt
\lower 8pt \hbox{\epsfxsize=160pt \epsfbox{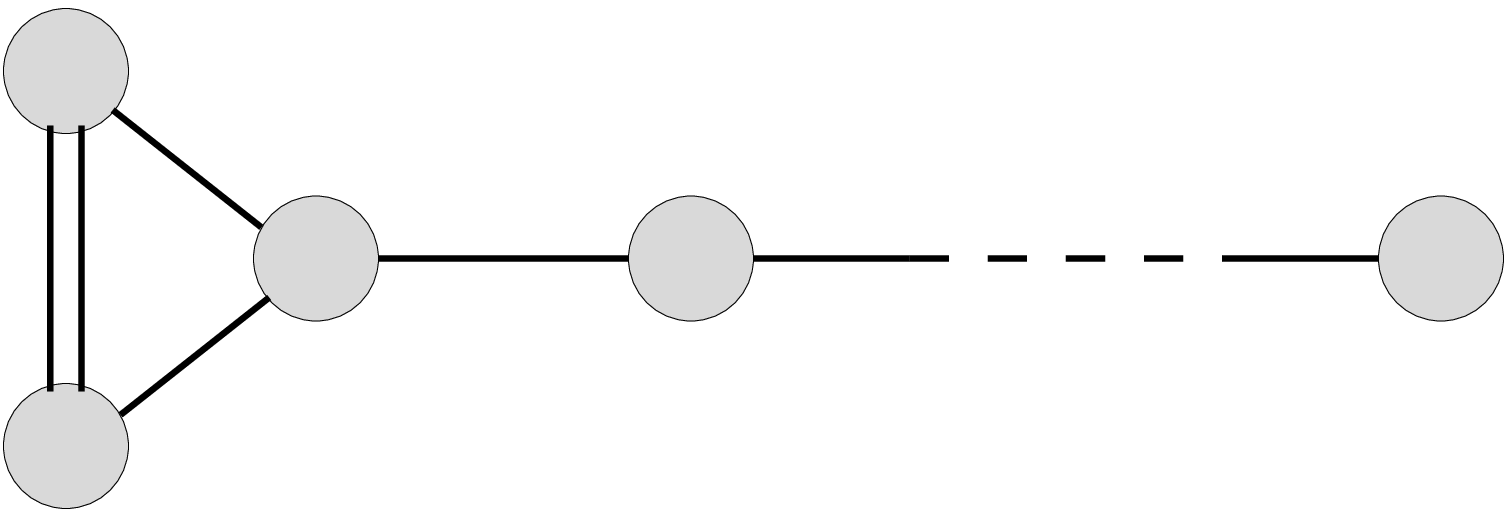}} }
\vskip 15pt
\centerline{\bf Figure 2: classical fermionic Dynkin diagrams}
\vskip 10pt
\noindent
All simple root systems 
for classical Lie superalgebras can be obtained 
by combining the bosonic diagrams in Figure 1 with the fermionic diagrams 
in Figure 2. 
Any Dynkin diagram for a finite-dimensional, classical superalgebra
consists of a chain of alternating blocks of type A and A* 
which may in addition have a tail of type B, B*, D, D*, or C at one
end only.
Any Dynkin diagram for an infinite-dimensional, classical superalgebra
consists of a chain of alternating blocks of type A and A* 
which either has its ends joined to form a circle, 
this being permissible if the total number of grey nodes
is even, or else has tails of type B, B*, D, D* or C attached at both ends.
Any such diagrams are allowed. 

We will make much further use of the idea that the diagrams in Figures 1
and 2 are building blocks for general Dynkin diagrams.
In particular, this suggests an obvious notation for 
various other classes of diagrams.
For instance, the Dynkin diagrams for the classical, bosonic affine algebras 
can be denoted:
\settabs 5\columns
\vskip 5pt
\+&\kern -8pt {\bf Algebra} && \kern -2pt {\bf Type of Diagram}\cr
\vskip 5pt
\+&$\a_{m-1}^{(1)}$ && $\widehat \a$ \ \ \ \ $m$ nodes \cr
\+&$\b_{m}^{(1)}$ && B-D \ $m{+}1$ nodes \cr
\vskip 3pt
\+&$\c_{m}^{(1)}$ && C-C \ $m{+}1$ nodes \cr
\vskip 3pt
\+&$\d_{m+1}^{(1)}$ && D-D \ $m{+}2$ nodes \cr
\+&$\a_{2m-2}^{(2)}$ && B-C \ $m$ nodes \cr
\+&$\a_{2m-1}^{(2)}$ && C-D \ $m{+}1$ nodes \cr
\+&$\d_{m+1}^{(2)}$ && B-B \ $m{+}1$ nodes \cr
\vskip 5pt
\noindent
where $m \geq 2$ 
(compare with Figures 3 and 5 of [\Ref{GO}] or Tables Aff1 and Aff2 of
[\Ref{Kac0}]). 
The symbol $\widehat \a$ indicates a diagram of type A 
whose ends have been joined
to form a circle, while the other symbols indicate a joining 
of two diagrams of type B, C or D. 
(Note that when there are just two nodes
$\widehat \a$ represents the same diagram as B-B or C-C; 
when there are three nodes 
B-D coincides with C-C, and C-D coincides with B-B; when there are
four nodes $\widehat \a$ coincides with D-D.)
We shall use a similar notation when we come to discuss fermionic diagrams,
and in preparation we introduce 
two further classes $\bl$ and $\dl$ which contain both 
bosonic and fermionic nodes, as shown in Figure 3.
Although these are composites of diagrams already defined, 
they will prove useful later. 
\vskip 10pt

\hbox{ \kern 100pt \hbox{{\br B}\lower5pt\hbox{\br *}} \kern 9pt
\lower 3pt \hbox{\epsfxsize=130pt \epsfbox{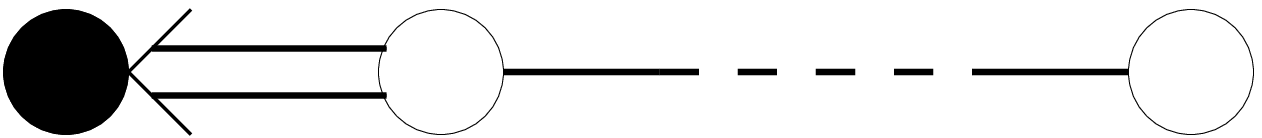}} } \vskip 20pt

\hbox{ \kern 100pt \raise 15pt \hbox{{\br D}\lower5pt\hbox{\br *}} \kern 12pt
\lower 8pt \hbox{\epsfxsize=120pt \epsfbox{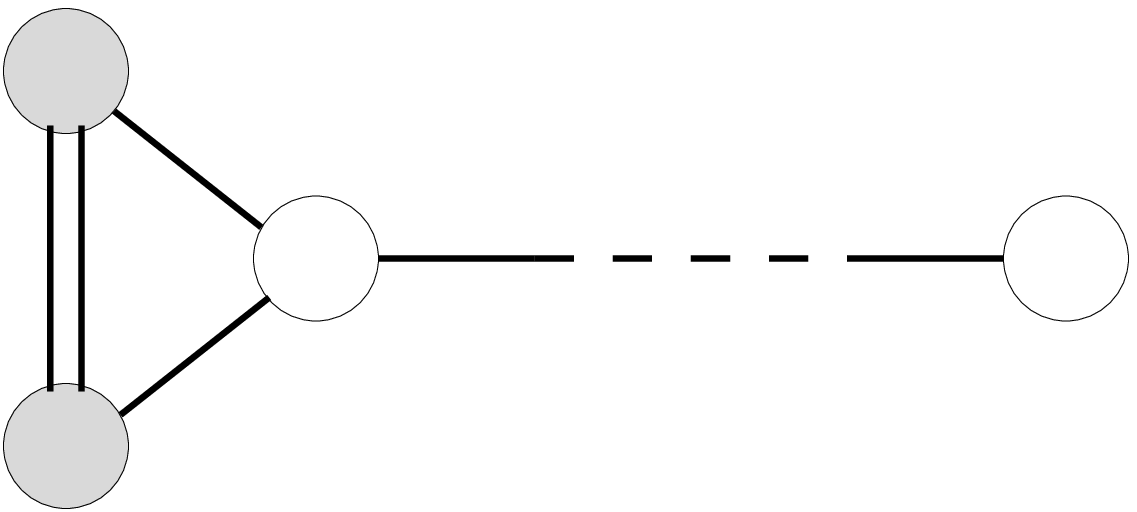}} }
\vskip 10pt

\centerline{\bf Figure 3: two additional families of classical Dynkin diagrams}
\vskip 10pt

We now supply the final ingredient in the correspondence between 
simple root systems and Dynkin diagrams for classical algebras
by describing how to reconstruct the signs of the entries in 
a Cartan matrix from a diagram.
For bosonic or fermionic simple roots with non-vanishing
length-squared
this is straightforward: in any connected part 
of a diagram which involves only white or black nodes, these
nodes all represent numbers with the same sign, and the bonds attached
to them all represent 
numbers with the opposite sign. 
For diagrams with null fermionic roots things are more 
complicated, however, 
and the relative signs must be taken in accordance with
those written in Figure 4. (The associated 
bosonic systems $\B_\pm$ will be introduced shortly.) 
The signs on the dotted bonds of the fermionic diagrams indicate the choices 
necessary
for an additional bond which appears in forming any larger, composite diagram.
Notice that these rules could not
be applied consistently to a circle containing an odd number of grey nodes;
but, as mentioned above, such diagrams are  
forbidden by the classification of superalgebras. 
\vskip 15pt
\hbox{ {\br A*} \kern 5pt 
\lower 3pt \hbox{\epsfxsize=165pt \epsfbox{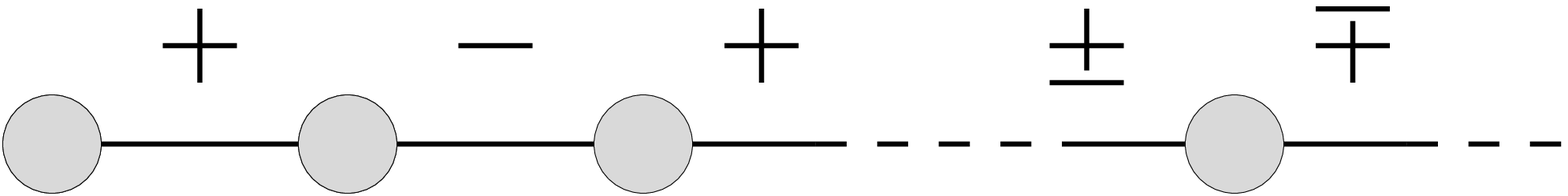}} \kern 18pt 
\lower 10pt \hbox{
\vbox{\hbox{\ca B$_+$} \vskip 10pt \hbox{\ca B$_-$}} }
\kern 3pt
\lower 12pt \hbox{{\epsfxsize=145pt \epsfbox{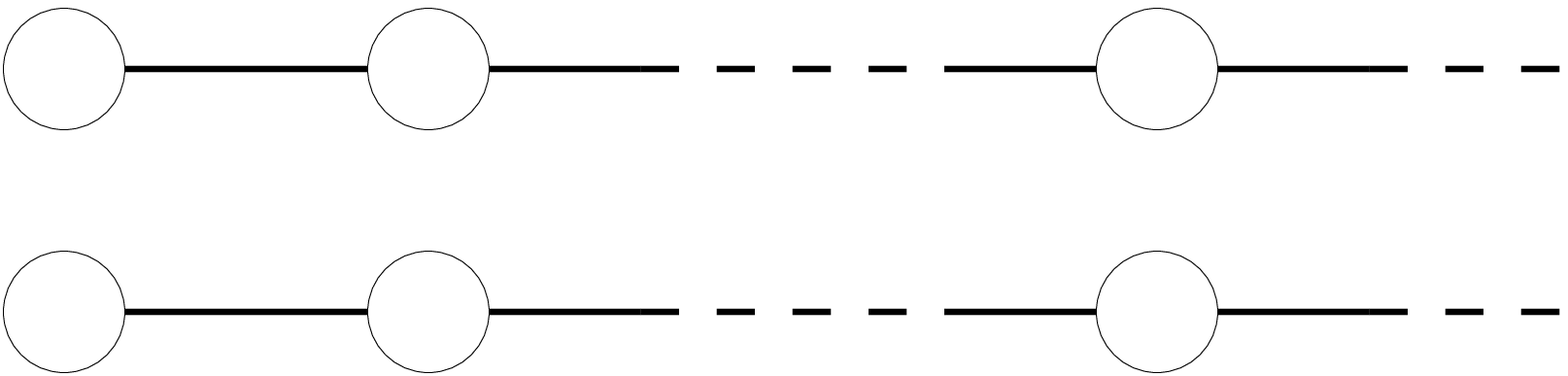}} 
}} 
\vskip 25pt

\hbox{ {\br B*} \kern 4pt
\lower3pt \hbox{\epsfxsize=170pt \epsfbox{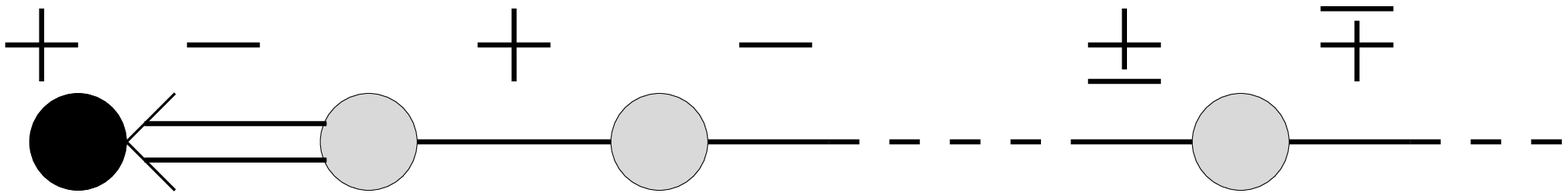}} \kern 18pt
\lower10pt \hbox{\vbox{ \hbox{\ca B$_+$} \vskip 10pt \hbox{\ca B$_-$}}}
\kern 5pt
\lower 12pt \hbox{{\epsfxsize=145pt \epsfbox{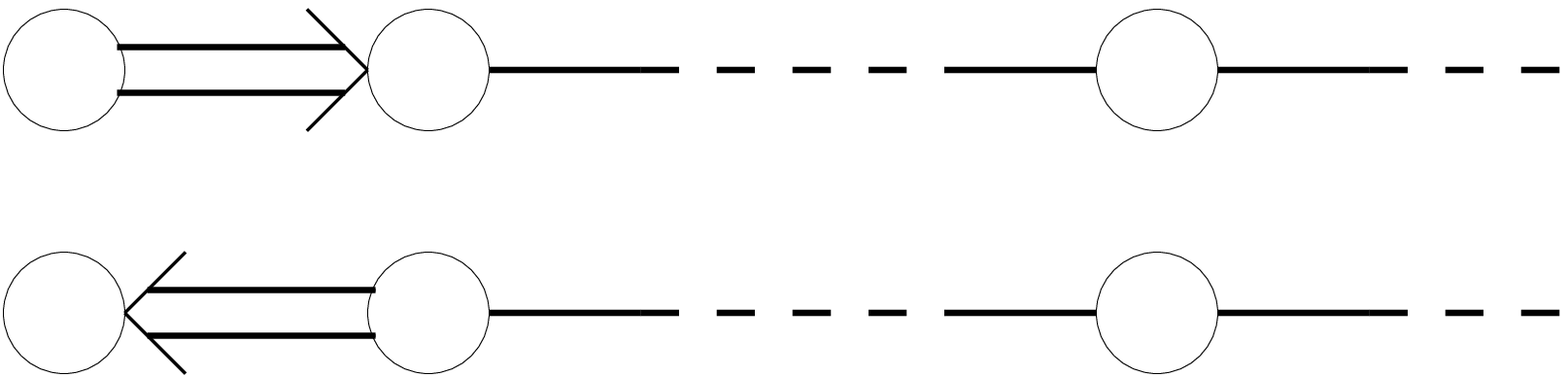}} 
}} 
\vskip 25pt

\hbox{ \raise 10pt \hbox{\br D*} \kern 5pt
\lower 8pt \hbox{\epsfxsize=170pt \epsfbox{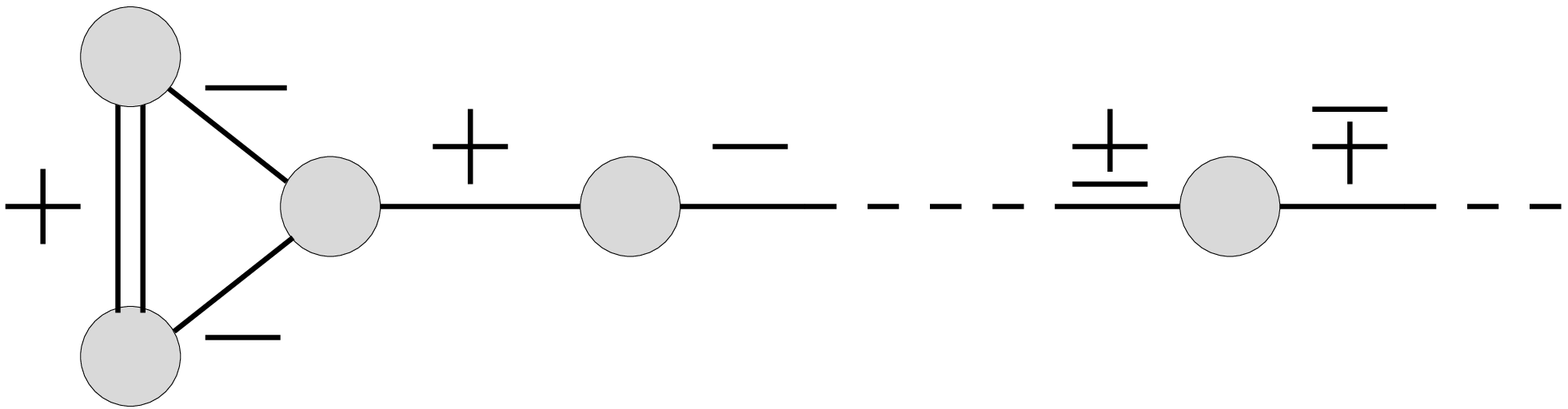}} \kern 18pt 
\vbox{\hbox{\ca B$_+$} \vskip 20pt \hbox{\ca B$_-$}}
\kern 5pt
\lower 20pt \hbox{{\epsfxsize=145pt \epsfbox{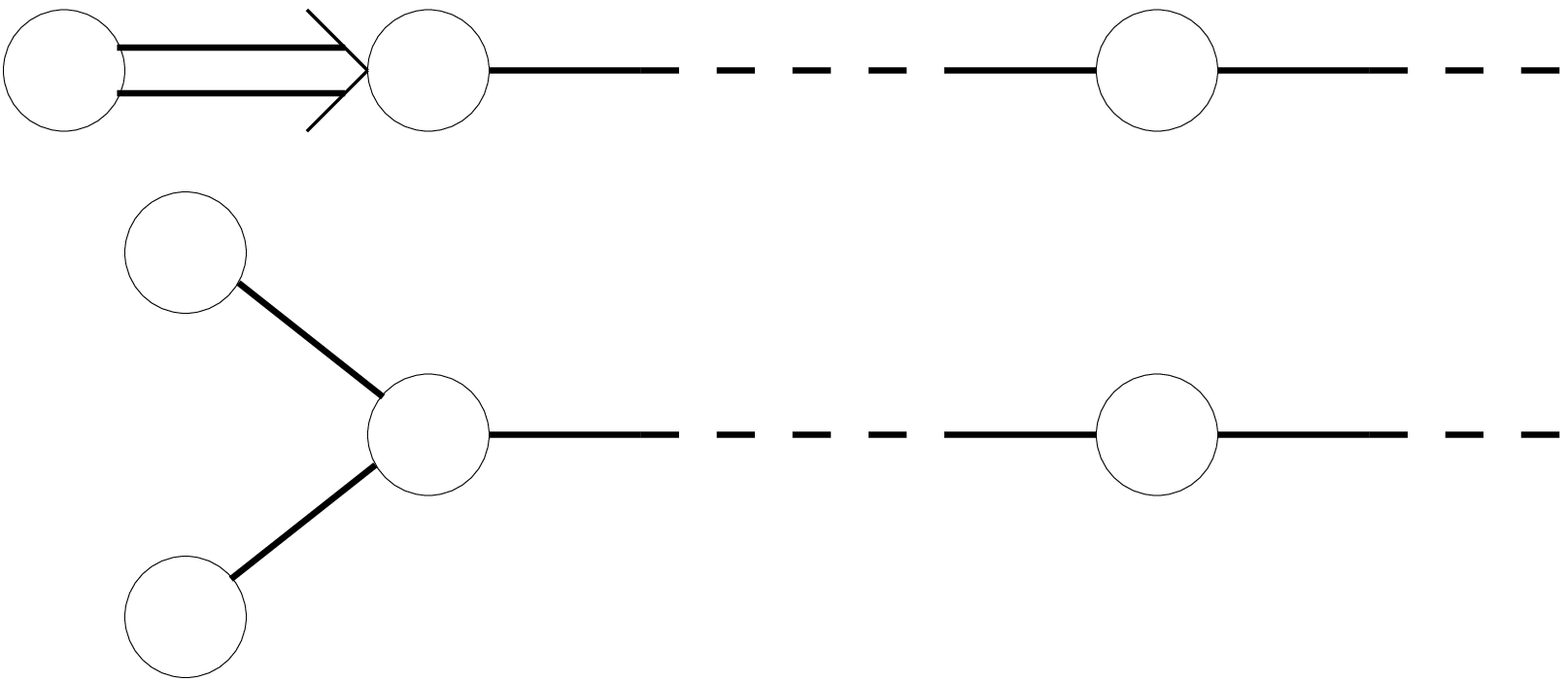}} 
}}
\vskip 15pt
\centerline{\bf Figure 4: signs associated with fermionic Dynkin diagrams}
\vskip 15pt 

It would be nice to give a set of unambiguous rules for reconstructing
such signs from any Dynkin diagram. Unfortunately, this is not
possible if one adheres to the standard way of drawing Dynkin 
diagrams---which is now more or less universally followed---because 
there is an inherent ambiguity. The difficulty arises with 
diagrams of the type shown in Figure 5, where the dotted bonds can be
of various of strengths. 
\vskip 15pt
\centerline{\epsfxsize=110pt\epsfbox{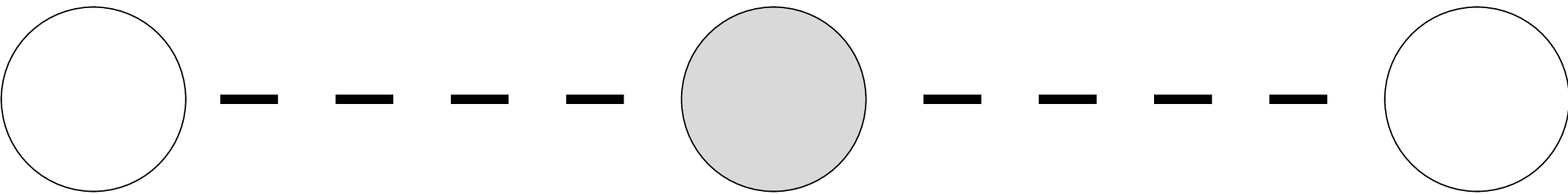}} 
\vskip 10pt
\centerline{\bf Figure 5: ambiguous Dynkin diagrams}
\vskip 10pt
\noindent
If these bonds are of strength one, for instance,
the diagram could represent root systems for either of the algebras
D(2,1) or A(1,1), depending on whether we take the signs corresponding
to the bonds to be equal or opposite.
This means that there is not quite
a one-to-one correspondence between diagrams and simple roots systems 
for superalgebras
(and this is quite distinct from the fact that many different simple root
systems can lead to isomorphic algebras).
In practice, however, this ambiguity is not very serious because it
involves essentially only this one particular class of diagrams.

It is because of this complication that 
we have chosen to present things initially in a rather concrete way.
We have also concentrated on the classical algebras
because we 
shall not have much reason to deal directly with the exceptional
algebras in this paper. Nevertheless, now that we have 
explained the difficulties, we can give a set of rules which is 
as clear-cut as possible and
which allows one to recover the signs 
corresponding to any Cartan matrix, including those for
exceptional algebras. 
These rules apply to {\it any Dynkin diagram\/} or {\it subdiagram\/}
(obtained by deleting a number of nodes and their attached bonds).
\vskip 2pt
\noindent 
{\bf $\bullet$} For any white or black node in any diagram, 
the sign on the node is opposite to the sign on any bond attached to it.
\vskip 2pt
\noindent 
{\bf $\bullet$} For any diagram without branches
(ie.~consisting of a line or a circle of nodes)
the signs on two bonds attached to any given grey node must be
opposite to one another, with the following exceptions:
\hfill \break
{\bf (i)} diagrams of the form shown
in Figure 5, for which the signs may be equal or opposite;
\hfill \break
{\bf (ii)} triangles of three grey nodes, which are dealt 
with by the next rule.
\hfill \break
(These exceptions involve members of the family
D(2,1;$\alpha$) for various values of $\alpha$.)
\vskip 2pt
\noindent 
{\bf $\bullet$} For any diagram consisting of a triangle of three grey
nodes, the signs on the bonds must be such that 
the sum of the strengths of the bonds, taken with 
these signs, is zero.
\vskip 2pt
\noindent 
It is easy to see that these rules hold for all the classical examples
given in Figure 4, and indeed that the signs given there are
necessary consequences of these rules. 
The exceptional algebras G(3), F(4) and their affine extensions can be 
considered similarly;
the Dynkin diagrams 
can be found in [\Ref{FSS}].

Now that we have clarified some aspects of the relationship between
Cartan matrices and Dynkin diagrams,
we consider a general 
simple root system $\R = \B \cup \F$ and introduce two associated 
bosonic simple root systems $\B_{\pm}$.
\vskip 2pt
\noindent 
{\bf Definition:} 
$\B_{\pm } = \{ \, \alpha_i \in \B \ | \ \alpha_i^2 \gle 0 \, \} \cup  
\{ \, \alpha_i + \alpha_j \, ; \, \alpha_i, \alpha_j \in \F \ | \ 
(\alpha_i + \alpha_j)^2 \gle 0 \, \}$.
\vskip 2pt
\noindent 
The sets $\B_{\pm}$ are clearly contained within the bosonic subsystem
of the entire root system of $\G(\R)$. If we 
consider the bosonic subalgebras $\G(\B_\pm)$ which are generated by $\B_\pm$,
these must be contained within $\G(\R)_{\rm bos}$.
The following observation will play a central role 
in all the following developments.
\vskip 5pt 
\noindent {\bf Proposition 1:} Let $\R$ be any superalgebra 
simple root system with
$\B_\pm$ defined as above.
\hfil \break
{\bf (a)} The sets $\B_\pm$ are
mutually orthogonal and the restrictions of the inner-product on 
$\langle \R \rangle $ to the subspaces $\langle \B_\pm \rangle $
are positive/negative-definite. 
Consequently $\G(\B_\pm)$ commute and we have a direct sum of 
Lie algebras $\G(\B_+) \oplus \G(\B_-) \subset \G(\R)_{\rm bos}$.
\hfil \break
{\bf (b)} If $i, j \in \tau$ then $\alpha_i +
\alpha_j \in \B_\pm$ iff $\alpha_i \cdot \alpha_j \gle 0$.
\vskip 2pt 
\noindent
Proof: (a) The fact that the algebras commute follows immediately
from the property that the sets of roots $\B_\pm$ are
mutually orthogonal, which can be established by inspection.
On a simple bosonic Lie algebra, the invariant inner-product is
essentially unique and must have definite signature.
Hence, we need only check that 
the inner-product is positive or negative on any one simple root in
each simple factor. It follows from the definitions 
of $\B_\pm$ that it must be positive/negative-definite on
$\langle \B_\pm \rangle$.
Part (b) is immediate if the roots are both null, but it is also easy to
check that it holds for the combination of a grey and a black node,
or two black nodes, which occur only in (sub)diagrams of type $\bu$. 
\vskip 5pt

The significance of the sets $\B_\pm$ will become clearer in the next
section. Notice that part (b) of Proposition 1 allows 
the sets of roots $\B_\pm$ to be read off very easily from the Dynkin
diagram for $\R$ with the pattern of relative $\pm$ signs attached 
according to
Figure 4. Since this observation proves so useful,    
we have recorded the results for $\B_\pm$ in Figure 4 for the primitive 
kinds of fermionic diagrams $\au$, $\bu$ and $\du$. 
These results can easily be combined to find $\B_\pm$ for more
complicated root systems $\R$.
In general the diagrams for $\B_\pm$ need not be connected
and, correspondingly, the algebras $\G(\B_\pm)$ may consist of direct sums 
of various simple factors.

%3
\chapter{Simple root systems and integrable models}

First we establish some conventions. 
Two-dimensional super-Minkowski space has bosonic 
light-cone coordinates $x^\pm$ and fermionic 
light-cone coordinates 
$\theta^\pm$ which are real (or Majorana) spinors. 
Lorentz transformations act on these coordinates according to
$x^\pm \rightarrow e^{\pm \lambda} x^\pm$
and $\theta^\pm \rightarrow e^{\pm \lambda/2} \theta^\pm $ 
where $\lambda$ is the rapidity.
Covariant super-derivatives obey 
$$
D_{\pm}= {\del \over \del \theta^{\pm}} -  i \theta^{\pm} \del_{\pm}
\, , \quad D^2_\pm = - i \del_\pm \ , \quad \{ D_+ , D_-\} = 0 \ .
$$
A general bosonic scalar superfield has a component expansion 
$$
\Phi=\phi + i \theta^+ \psi_{+}  + i \theta^- \psi_{-} + 
i \theta^+ \theta^- \sigma \ .
$$
Note that the fermionic fields $\psi_\pm$ are components of a spinor 
which transform under the Lorentz group according to
$\psi_\pm \rightarrow e^{\mp \lambda/2} \psi_\pm $, 
in contrast to the behaviour of the fermionic coordinates.
This is consistent because 
spinor indices can be raised and lowered by reversing signs,
though we shall never need to make use of this. We shall 
write spinor indices explicitly so as to avoid the introduction of 
Dirac matrices. However, we will adopt the usual notation for
the Lorentz-invariant and parity-even inner-product
$$
\chib \psi = i (\chi_+ \psi_- + \psi_+ \chi_- ) \ ,
$$
which is real and manifestly symmetric for real fermions.

Now given any simple root system of a Lie superalgebra 
$\R = \B \cup \F$
we can write down a corresponding integrable Toda theory
in terms of a real scalar superfield $\Phi$ which takes 
values in the target space $V = \langle \R \rangle$.
The equations of motion are
$$
iD_+  D_- \Phi = \sum_{j \in \tau}
\m_j \alpha_j\exp{\alpha_j\cdot\Phi}
+ i \theta^+ \theta^- \sum_{j \in \varepsilon}
\m_j \alpha_j\exp{\alpha_j\cdot\Phi}
\nfr{eqm}
and they can be derived from the superspace lagrangian density
$$
L={i\over2} D_+ \Phi \cdot D_- \Phi 
+ \sum_{j \in \tau}\m_j\exp\alpha_j\cdot\Phi
+ i \theta^+ \theta^- \sum_{j \in \varepsilon} \mu_j 
\exp \alpha_j \cdot \Phi \ .
\nfr{slag}
For discussions of integrability of these equations see
[\Ref{LS}--\Ref{HR},\Ref{O}--\Ref{EH1},\Ref{NM}--\Ref{IK},\Ref{QI}].
The quantities $\m_i$ are non-zero parameters with the dimensions of
({\it mass\/}) for $i \in \tau$ and ({\it mass\/})${}^2$ for 
$i \in \varepsilon$. 
These parameters can clearly be modified to some extent by shifting the 
fields $\Phi$, but it is convenient to keep their values arbitrary 
for the moment. 
If desired one can introduce a dimensionless coupling $\beta$ 
by taking $L \rightarrow L/\beta^2$, and often 
the Toda equations are written in a way which corresponds to
taking $\Phi \rightarrow \beta \Phi$ in addition. This coupling
will not be very important for us, since we work entirely classically.  

To pass to the component field formulation of 
the model, we expand the exponential terms and eliminate the auxiliary
fields $\sigma$ in the standard fashion to obtain 
$$
L = L_{\rm bos} + L_{\rm ferm} + L_{\rm int} ,
\efr
where the terms appearing in the component lagrangian are
$$
\eqalign{
L_{\rm bos} & = {1 \over 2} \, \del_+ \phi \cdot \del_- \phi 
- \sum_{j \in \varepsilon}\m_j\exp\alpha_j\cdot\phi
- \sum_{i,j \in \tau} {1 \over 2}
\m_i\m_j \, (\alpha_i \cdot \alpha_j ) 
\exp (\alpha_i + \alpha_j) \cdot \phi
\cr
L_{\rm ferm} & = {i \over 2} \, \psi_+ \cdot \del_- \psi_+
+ {i \over 2} \, \psi_- \cdot \del_+ \psi_-
\cr
L_{\rm int} & = \sum_{j \in \tau} i \m_j (\alpha_j \cdot \psi_+) 
(\alpha_j \cdot \psi_-) \exp \alpha_j \cdot \phi \ .
\cr
}
\nfr{clag}
For future reference we also record the component field equations: 
$$\eqalign{
\del_+ \del_- \phi =
&-\sum_{j \in \varepsilon}\m_j 
\alpha_j \exp\alpha_j\cdot\phi
- \sum_{i,j \in \tau} {1 \over 2} 
\m_i\m_j \, (\alpha_i + \alpha_j) (\alpha_i \cdot \alpha_j) \exp 
(\alpha_i + \alpha_j) \cdot \phi
\cr
&\quad + \sum_{j \in \tau} i \m_j \alpha_j 
(\alpha_j \cdot \psi_+) (\alpha_j \cdot \psi_-)
\exp \alpha_j \cdot \phi \ ,
\cr
\del_\mp\psi_\pm = 
& \mp \sum_{j \in \tau} \mu_j \alpha_j (\alpha_j \cdot \psi_\mp)
\exp \alpha_j \cdot \phi \ .
\cr
}
\nfr{cem}
It should be clear either from the superspace description or from the
component version written above that the Toda model really does depend 
on the particular simple root system $\R$ rather than on the 
superalgebra $\G (\R)$. We will see examples
below of Toda models with very different properties which are
based on inequivalent simple root systems for the same superalgebra.

The bosonic sector, 
with lagrangian $L_{\rm bos}$ in \clag , 
must be integrable by itself,
because if the fermions are set to zero at some initial time then they
remain zero according to the full set of equations of motion. 
But it is clear that this bosonic model is not necessarily a Toda
theory based on a {\it simple\/} Lie algebra. 
Rather, $L_{\rm bos}$ will generally consist of 
a sum of several independent bosonic Toda theories, 
possibly including a number of massless free fields. 
Each of these independent factors may contribute to the 
lagrangian with either a positive or a negative sign.  
It is therefore appropriate at this stage to try to clarify some troublesome 
points concerning the positivity of actions, kinetic energy, and potential
energy. 

It is very important to distinguish between the {\it overall\/}
sign of the lagrangian, which is irrelevant at the classical level,
and the {\it relative\/} sign between the kinetic and
potential terms, which has much more immediate physical significance.
This relative sign determines whether a given classical solution is
stable, and in particular the sign must be chosen correctly if 
the theory is to have oscillations with positive $(mass)^2$ about
a given stationary point of the potential.
This is why we consider a bosonic lagrangian 
$\Half \del_+ \phi \cdot \del_- \phi - U(\phi)$ to be physically sensible
when the inner-product in the kinetic term is positive-definite and
the potential $U(\phi)$ is bounded below. We shall say that any 
lagrangian or action of this type is {\it positive-definite\/}, or that it has
{\it positive signature}.
But from the classical point of view we could equally well
change the lagrangian by an overall sign, or equivalently take the
inner-product to be negative-definite and the potential to be bounded
above. We refer to such a lagrangian or its action 
as being {\it negative-definite\/} or 
having {\it negative signature}.
If the inner-product is neither positive- nor negative-definite we
shall say that the lagrangian has {\it indefinite signature\/}.

We must expect $L_{\rm bos}$ in \clag \ to have 
indefinite signature, in general. 
The corresponding Toda equations \cem \ will still be 
integrable (in the 
sense discussed in [\Ref{O}-\Ref{QI}]) no matter what values of the
parameters $\mu_i$ we choose, but when we
pay closer attention to the relative signs
between kinetic and potential terms in \clag , we
find that there are some choices of particular 
physical interest. In fact, we can always ensure that $L_{\rm bos}$ 
is a combination of bosonic Toda models which are all {\it individually\/}
positive- or negative-definite, even though their sum may have
indefinite signature. 
\vskip 5pt

\noindent
{\bf Proposition 2:} Consider the Toda theory based on a simple root system
$\R$. The bosonic sector has a lagrangian of the form 
$$
L_{\rm bos} = L_+ + L_- + L_0 
\nfr{bos} 
where $L_{\pm}$ are lagrangians for bosonic Toda theories 
based on simple root systems $\B_\pm$ and $L_0$ is the lagrangian
for a number of free massless scalars fields. 
This sum of lagrangians corresponds to an orthogonal decomposition of
the target space
$V = V_+ \oplus V_- \oplus V_0$ where 
$V_{\pm} = \langle \B_\pm \rangle $ are the target spaces for 
$L_\pm$ and $V_0 = ( \langle \B_+ \rangle \oplus \langle \B_- \rangle
)^\bot $ is the target space for $L_0$. 
The lagrangians $L_\pm$ are 
positive/negative-definite if we choose
$\mu_i \gle 0$ for $\alpha_i^2 \gle 0$ when $i \in \varepsilon$,
and $\mu_i$ all with the same sign when $i \in \tau$.
Each of the scalar fields in $L_0$ may contribute with 
a positive or negative sign. 
\vskip 2pt
\noindent
Proof: These statements can all be checked directly from \clag \ in 
conjunction with the results of Proposition 1. In particular, 
part (b) of Proposition 1 ensures the positive/negative-definite
nature of $L_\pm$ with the choices of signs for $\mu_i$ given above.
\vskip 5pt

This gives an essentially complete description of 
the bosonic sector of any Toda theory based on a superalgebra.
Moreover, the relevant semi-simple algebras with root systems 
$\B_\pm$ can be determined very easily for any given $\R$ 
by using part (b) of Proposition 1 together
with the information summarized in Figure 4.
Taken together, these facts constitute a simple diagrammatic
method for determining the bosonic sector of any Toda theory,
and much of what follows in the rest of this paper rests on this idea.
Some examples will be given in the following sections. 

When we turn to consider the fermions in \clag \ we see that 
$L_{\rm ferm}$ is a free lagrangian, and that all the remaining 
non-trivial interactions are contained in $L_{\rm int}$.
We should emphasize that these interactions will 
link together the various components 
of $L_{\rm bos}$ described in Proposition 2; it is only when the
fermions vanish that these components decouple.
We will not attempt to analyze the structure of $L_{\rm int}$ 
in any generality, except for the following observation. 
\vskip 5pt

\noindent
{\bf Proposition 3:} Consider the Toda theory based on a simple root
system $\R$. If the inner-product on $V = \langle \R \rangle $
is non-degenerate when
restricted to the subspace $\langle \F \rangle $ spanned by the fermionic
simple roots, then 
$V = \langle \F \rangle \oplus \langle \F \rangle^\bot$.
In this case the only relevant fermionic degrees of freedom are those
fields in $\langle \F \rangle $, while those in $\langle \F \rangle^\bot $
obey free, massless equations of motion and are decoupled from all other
fields. 
\vskip 2pt
\noindent
Proof: First notice that if $\omega$ is any vector in $\langle \F
\rangle^\bot$ then the fermion field 
$\omega \cdot \psi_\pm$ always obeys 
a free equation of motion according to \cem . 
To show that a given field decouples, however, we must also demonstrate
that it does not act as a source for other fields by being present in
the interaction terms.
From the expression for $L_{\rm int}$, it is clear that 
the fermion fields which are absent from the interaction terms are
precisely those living in the subspace $\langle \F \rangle^\bot$.
When the inner-product is non-degenerate on 
$\langle \F \rangle$ we can decompose any $\psi$ as a sum of 
fields in $\langle \F \rangle$ and $\langle \F \rangle^\bot$
in a unique way, and the result follows.
\vskip 5pt

\noindent
{\bf Corollary:} If there are no fermionic null roots in $\R$, the 
fermions in $\langle \F \rangle^\bot$ always decouple,
since the inner-product is positive-definite on $V = \langle \R
\rangle$ and so non-degenerate on {\it any\/}
subspace (this is the situation discussed in [\Ref{O},\Ref{WI}]).
An even more special case arises when there are no fermionic roots at
all, so $\R = \B$.
Then all the fermions are decoupled, $L_{\rm bos}$ is 
the usual bosonic Toda theory based on $\B$ and is the only
non-trivial part of \clag .
In this sense the superspace construction
contains the usual bosonic Toda construction [\Ref{LS}-\Ref{OT}] 
as a special case.
\vskip 5pt

We now summarize the classification of two important categories of
Toda models.
It turns out that they share the rather special property 
that their bosonic sectors correspond to the {\it entire\/} bosonic
subalgebras $\G (\R)_{\rm bos}$ of $\G (\R)$.
\vskip 5pt

\noindent
{\bf Proposition 4:} Consider the Toda theory based on a simple root
system $\R$. 
\hfil \break
{\bf (a)} The theory is supersymmetric
iff the simple root system is entirely fermionic, $\R = \F$; 
a complete list is given in Table 1 
(following [\Ref{LSS},\Ref{EH1},\Ref{EH2}]). 
\hfil \break
{\bf (b)} The theory with real fields has a positive-definite lagrangian 
iff $\R$ has no null roots; 
a complete list of such systems involving at least one fermionic root is
given in Table 2 (following [\Ref{Kac2},\Ref{O}]). 
\hfil \break
{\bf (c)} In each of these cases we
have $\G ( \B_ + ) \oplus \G ( \B_- ) = \G (\R )_{\rm bos}$, whereas in
general the left-hand side can be a proper subalgebra. 
\vskip 15pt

{\bf Table 1: Totally fermionic simple root systems}
\vskip 10pt

\settabs\+XX&XXXX&SuperalgebraXXXXXX&Simple&Xroot
systemXXXXX&Bosonic subalgebra\cr

{\bf (a) Linearly independent systems for finite-dimensional superalgebras } 
\vskip 2pt

\+ &&{\bf Superalgebra} & {\bf Simple root system} && 
{\bf Bosonic subalgebra} \cr
\vskip 2pt
\+&& $\a(m, m{-}1)$ & $\au$ \ &$2m$ nodes & $\a_m\oplus\a_{m-1}\oplus\u(1)$ \cr
\+&& $\b(m{-}1,m)$  & $\bu$ \ &$2m{-}1$ nodes & $\b_{m-1} \oplus \c_m $ \cr
\+&& $\b(m,m)$  & $\bu$ \ &$2m$ nodes & $\b_m \oplus \c_m$ \cr
\+&& $\d(m{+}1,m)$ & $\du$ \ &$2m{+}1$ nodes & $\d_{m+1} \oplus \c_m$ \cr
\+&& $\d(m,m)$  & $\du$ \ &$2m$ nodes & $\d_m \oplus \c_m$\cr
\+&& $\d(2,1; \alpha)$ & &--- & $\a_1 \oplus \a_1 \oplus \a_1$ \cr
\vskip 10pt

{\bf (b) Linearly dependent systems for finite-dimensional superalgebras } 
\vskip 2pt
\+ &&{\bf Superalgebra} & {\bf Simple root system} && 
{\bf Bosonic subalgebra} \cr
\vskip 2pt
\+&& $\a(m,m)$ & $\au$ \ &$2m{+}1$ nodes & $\a_m \oplus \a_m $ \cr
\vskip 10pt

{\bf (c) Linearly dependent systems for infinite-dimensional superalgebras } 
\vskip 2pt
\+&&{\bf Superalgebra}&{\bf Simple root system} && {\bf Bosonic subalgebra} \cr
\vskip 2pt
\+&& $\a(m,m)^{(1)}$ & $\widehat \au$ &\ \ $2m{+}2$ nodes 
& $\a_m^{(1)} \oplus \a_m^{(1)}$ \cr
\+&& $\a(2m,2m)^{(4)}$ & B*-B*  &\ \ $2m{+}1$ nodes 
& $\a_{2m}^{(2)} \oplus \a_{2m}^{(2)}$ \cr
\+&& $\a(2m{+}1,2m{+}1)^{(2)}$ & D*-D* &\ \ $2m{+}3$ nodes 
& $\a_{2m+1}^{(2)} \oplus \a_{2m+1}^{(2)}$ \cr
\+&& $\a(2m{+}1,2m)^{(2)}$ & B*-D* &\ \ $2m{+}2$ nodes
& $\a_{2m+1}^{(2)} \oplus \a_{2m}^{(2)} $ \cr 
\+&& $\b(m,m)^{(1)}$ & B*-D* &\ \ $2m{+}1$ nodes 
& $\b_m^{(1)} \oplus \c_m^{(1)}$ \cr
\+&& $\d(m{+}1,m)^{(1)}$ & D*-D* &\ \ $2m{+}2$ nodes 
& $\d_{m+1}^{(1)} \oplus \c_{m}^{(1)}$ \cr
\+&& $\d(m,m)^{(2)}$ & B*-B* &\ \ $2m$ nodes 
& $\d_m^{(2)} \oplus \c_m^{(1)}$ \cr
\+&& $\d(2,1;\alpha)^{(1)}$ &&\ \ --- 
& $ \a_1^{(1)} \oplus \a_1^{(1)} \oplus\a_1^{(1)} $ \cr
\vskip 10pt

\noindent
For all entries $m \geq 1$, where we define 
$\d(1,m) = \c(m{+}1)$ and note $\d(1,1) = \c(2) = \a(1,0)$.
Each of the superalgebras listed has a unique 
purely-fermionic simple root system which we have expressed in 
terms of the building blocks introduced in Figure 2. 
The notation $\widehat \au$ stands for a diagram consisting of a circle of
grey nodes, and the other symbols indicate diagrams 
formed by joining building blocks of types B* and D*,
as discussed in the last section.
Dynkin diagrams for the exceptional D families may be found in 
[\Ref{Dict}].
\vskip 15pt

{\bf Table 2: Simple root systems with positive-definite inner-product } 
\vskip 10pt

{\bf (a) Linearly independent systems for finite-dimensional superalgebras } 
\vskip 2pt
\settabs\+XX&No.XXXX&SuperalgebraXXXXXX&Simple& root
systemXXXXX&Bosonic subalgebra\cr
\+ &&{\bf Superalgebra} & {\bf Simple root system} && 
{\bf Bosonic subalgebra} \cr
\vskip 2pt
\+&& $\b(0,m)$    & $\bl$ \ \  &\ \ $m$ nodes &\ \ $\c_m $ \cr
\vskip 10pt

{\bf (b) Linearly dependent systems for infinite-dimensional superalgebras } 
\vskip 2pt
\+ &&{\bf Superalgebra} & {\bf Simple root system} && 
{\bf Bosonic subalgebra} \cr
\vskip 2pt

\+&& $\b(0,m)^{(1)}$ & $\bl$-C &\ \ \ $m{+}1$ nodes &\ \ $\c_{m}^{(1)} $ \cr
\+&& $\a(2m{+}1,0)^{(2)}$& $\bl$-D  &\ \ \ $m{+}2$ nodes &\ \
$\a_{2m+1}^{(2)}$\cr
\+&&$ \a(2m,0)^{(4)}$  & $\bl$-B  &\ \ \ $m{+}1$ nodes  &\ \
$\a_{2m}^{(2)}$\cr
\+&&$\c(m{+}1)^{(2)} $ & $\bl$-$\bl$ &\ \ \ $m{+}1$ nodes &\ \ 
$\c_{m}^{(1)} $ \cr
\vskip 10pt

\noindent
For all entries $m \geq 1$, 
and the building blocks signifying the type of diagram
include $\bl$ introduced in Figure 3. 
(The bosonic sectors given in Table 1 of [\Ref{O}] are incorrect in
some instances; this was also noted in [\Ref{STS}].)

We explained above that the overall sign of a lagrangian is 
essentially unimportant for a classical theory. 
But the situation is rather different at the quantum level:
the definition and interpretation of a quantum theory becomes
considerably more subtle if the lagrangian is not positive-definite.
Now on the one hand we have grown accustomed in recent years to the idea 
that indefinite or complex lagrangians can be useful 
quantum-mechanically in the description of non-unitary models,
or even of unitary models if some 
projection onto a subspace of physical states can be implemented
[\Ref{SGpert},\Ref{BRST},\Ref{TH}].
Having said that, it is also natural to retain a particular interest 
in those actions which are positive-definite---like those listed in
Table 2---which can be quantized in an entirely straightforward way.
In this paper we will certainly keep an open mind regarding 
the likely importance of theories with indefinite signature;
but we will also single out for special attention those theories
which are positive-definite. 

\chapter{Examples with extended conformal symmetry} 

Toda theories provide a valuable source of examples of  
extended conformal symmetries, in which the Virasoro algebra is
embedded within some larger ${\cal W}$-algebra. 
In fact the Toda construction, or the related procedure of Hamiltonian
reduction, provides one of the most systematic ways of
constructing ${\cal W}$-algebras, and may perhaps lead to some
eventual classification of them. 
The ${\cal W}$-algebras arising from bosonic Toda theories
have been investigated from a variety of different points of view.
The ${\cal W}$-algebras arising from superalgebra models are much less
well studied. In this context our results concerning the structure of
a general Toda model can provide some basic information on the bosonic 
sectors of these new extended conformal symmetries.

Let us now briefly recall how conformal symmetry arises. 
The following statements are well-known but it is useful to collect
them in a proposition for completeness and for future reference.
\vskip 5pt

\noindent
{\bf Proposition 5:} 
Given a simple root system $\R$ which is linearly independent,
the corresponding Toda model is invariant under 
conformal transformations $x^\pm \rightarrow y^\pm (x^\pm)$ acting on 
the fields according to 
$$\eqalign{
\phi (x^+ , x^-) & \rightarrow \phi (y^+,y^-) + 
\rho \log (\del_+ y^+ \del_- y^-) \cr
\psi_\pm (x^+ , x^-) &\rightarrow (\del_\pm y^\pm )^{1/2} 
\psi_\pm (y^+,y^-) \cr
}
\nfr{conf}
where the vector $\rho$ is defined by the conditions 
$$
\rho \cdot \alpha_i = \left \{ \matrix{ 1 &\ \ \ i \in \varepsilon\cr
1/2 & \ \ \ i \in \tau } \right .
$$
Linear independence of $\R$ (or $n=r$) ensures that $\rho$ exists;
this holds for all finite-dimensional Lie superalgebras,
except for the family $\a(m,m)$ (which is unusual in having $n = r{+}1$).
If $\R = \F$, so that the model is also supersymmetric, then 
conformal invariance is extended to superconformal invariance.
A complete list of superconformal models is therefore 
given in Table 1, section (a). 
\vskip 2pt
\noindent
Proof: These statements can be verified directly from \clag \ or \cem ;
for details see eg.~[\Ref{EH1}].
\vskip 5pt

A consequence of $\R$ being linearly independent is that 
we can always re-define the Toda fields by additive constants so as to
make $| \mu_i | = 1$. The fact that these mass parameters can essentially
be eliminated is a manifestation of conformal invariance.
The inhomogeneous pieces in the transformation of the bosons in
\conf\ give rise to a
classical central charge $12 \rho^2$ but this is generally
modified quantum-mechanically---see [\Ref{PM},\Ref{EH1}] and 
references given there.

To illustrate the results of the previous section, 
we list in Table 3 all linearly independent simple root systems of 
rank 1 or 2, together with a selection of root systems of rank
3. In each case we specify $\R$ by its Dynkin
diagram and give the corresponding bosonic systems $\B_\pm$ as found by
our diagrammatic method. The number of free scalar fields in the
bosonic sector of the Toda theory is indicated by the dimension of
$V_0$, and in all these examples they contribute with a negative sign.
Simple root systems of higher rank 
can be considered in an exactly similar fashion. 

The examples labeled (1)-(6) are particularly instructive 
and so we shall examine them more closely. 
We adopt the notation 
$\Phi_i = \alpha_i \cdot \Phi$ for superfields and their components. 
It will be convenient to denote the lagrangians for a free massless
boson $\phi$ and fermion $\psi$ by 
$$
L_{\rm free} (\phi) = \del_+ \phi \, \del_- \phi \ , \qquad
L_{\rm free} (\psi) = i \psi_+ \del_- \psi_+ + i \psi_- \del_+ \psi_-
\ ,
\nfr{free} 
respectively. The overall 
normalizations for various lagrangians appearing below will 
not always be standard, but we choose
them in order to eliminate as 
many awkward numerical factors as possible.
\vskip 20pt 

%%%%%%%%%%%%%%%%%% TABLE 3
\noindent{\bf Table 3: Some low-rank linearly-independent simple root
systems} 
\vskip 8pt
\settabs\+NoXXX&NameXXXXXX&RXXXXXXXXXXX&B+XXXXXXXX&B-XXXXXXXX&V0\cr
\+\phantom{XX} {\bf Superalgebra} && \phantom{XX} 
$\R$ & $\B_+$ & $\B_-$ &\kern -8pt dim$V_0$ \cr
\vskip 8pt
\+{\bf (1)} &$\a_1$
&\hbox{\epsfxsize=8pt\epsfbox{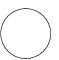}} 
&\hbox{\epsfxsize=8pt\epsfbox{p1.eps}}            
&---
&0 \cr
\vskip 5pt
\+{\bf (2)} &B(0,1) 
&\hbox{\epsfxsize=8pt\epsfbox{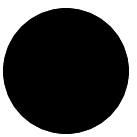}} 
&\hbox{\epsfxsize=8pt\epsfbox{p1.eps}}            
&---
&0 \cr
\vskip 5pt
\+{\bf (3)} &B(0,2) 
&\def\box1{\hbox{\epsfxsize=32pt\epsfbox{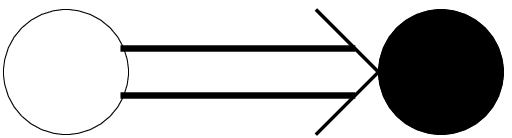}}}
\kern 28.5pt \rotf1   
&\def\box1{\hbox{\epsfxsize=32pt\epsfbox{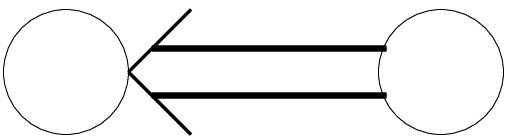}}}
\kern 28pt \rotf1   
&---
&0 \cr
\vskip 5pt
\+{\bf (4)} &A(1,0)
&\hbox{\epsfxsize=32pt\epsfbox{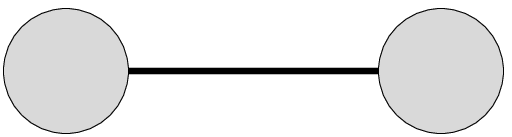}} 
&\hbox{\epsfxsize=8pt\epsfbox{p1.eps}} 
&---
&1 \cr
\vskip 5pt
\+{\bf (5)} &A(1,0) 
&\hbox{\epsfxsize=32pt\epsfbox{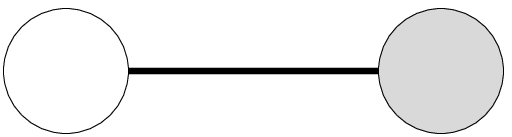}} 
&\hbox{\epsfxsize=8pt\epsfbox{p1.eps}} 
&---
&1\cr
\vskip 5pt
\+           &B(1,1) 
&\hbox{\epsfxsize=32pt\epsfbox{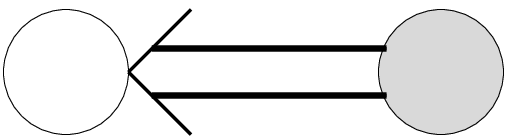}} 
&\hbox{\epsfxsize=8pt\epsfbox{p1.eps}} 
&---
&1 \cr
\vskip 5pt
\+           &B(1,1) 
&\hbox{\epsfxsize=32pt\epsfbox{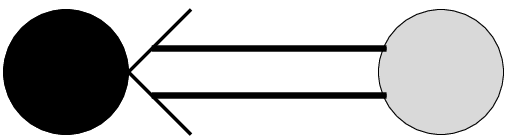}} 
&\hbox{\epsfxsize=8pt\epsfbox{p1.eps}} 
&\kern 1.5pt \hbox{\epsfxsize=8pt\epsfbox{p1.eps}} 
&0\cr
\vskip 5pt
\+           &A(2,0) 
&\hbox{\epsfxsize=56pt\epsfbox{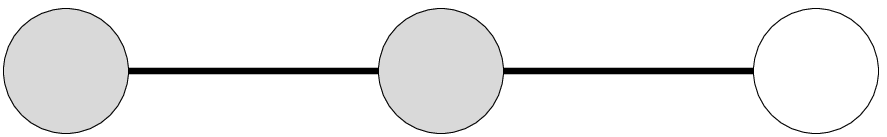}} 
&\hbox{\epsfxsize=32pt\epsfbox{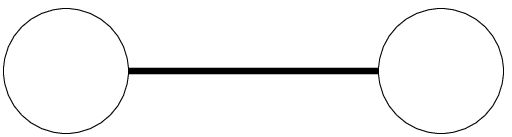}} 
&--- 
&1 \cr
\vskip 5pt
\+           &B(1,2)  
&\def\box1{
\hbox{\epsfxsize=56pt\epsfbox{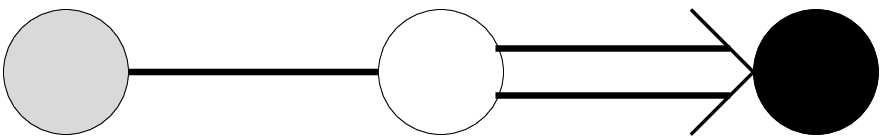}} 
} 
\kern 56.3pt \rotf1
&\def\box1{\hbox{\epsfxsize=32pt\epsfbox{B2.eps}}}
\kern 28.5pt \rotf1   
&--- 
&1 \cr
\vskip 5pt
\+           &B(2,1)
&\def\box1{
\hbox{\epsfxsize=56pt\epsfbox{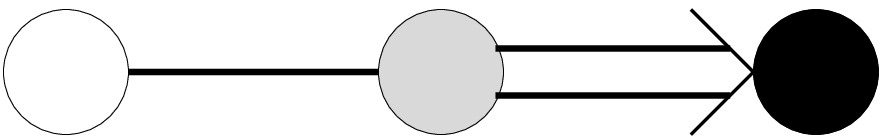}} 
}
\kern 56.3pt \rotf1
&\hbox{\epsfxsize=32pt\epsfbox{B2.eps}} 
&\kern 1.5pt \hbox{\epsfxsize=8pt\epsfbox{p1.eps}} 
&0\cr
\vskip 5pt
\+{\bf (6)}  &C(3) 
&\def\box1{\hbox{\epsfxsize=32pt\epsfbox{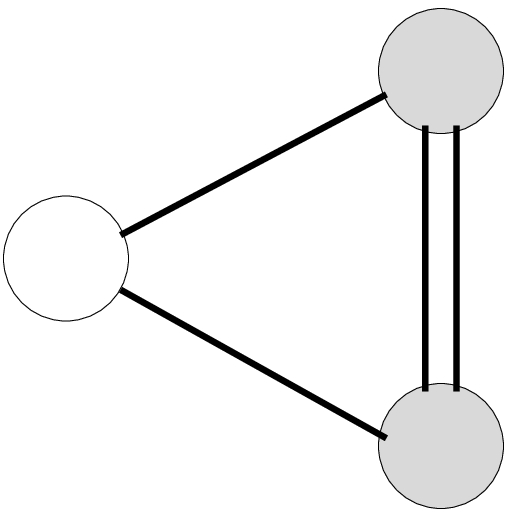}}}
\lower 12pt \hbox{\kern30pt \rotf1} 
&\def\box2{\hbox{\epsfxsize=32pt\epsfbox{B2.eps}}}
\hbox{\kern28.5pt \rotf2}   
&---
& 1\cr
\vskip 5pt
\+&D(2,1) 
&\def\box1{\hbox{\epsfxsize=32pt\epsfbox{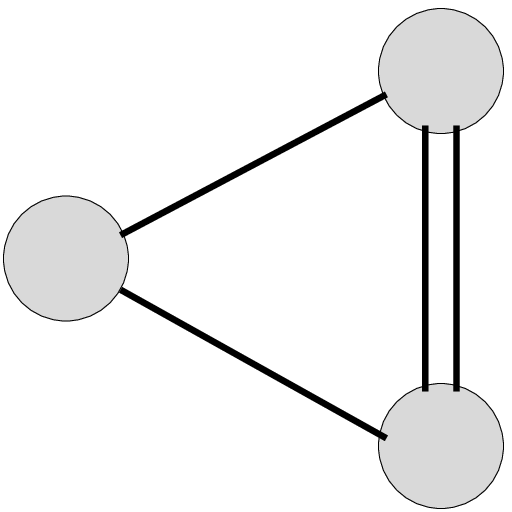}}}
\lower 12pt \hbox{\kern30pt \rotf1} 
&\hbox{\epsfxsize=32pt\epsfbox{A2.eps}}
&\kern 1.5pt \hbox{\epsfxsize=8pt\epsfbox{p1.eps}} 
&0\cr
%%%%%%%%%%%%%%%%%%%% END TABLE 3
\vskip 20pt

\noindent
{\bf (1)} $\a_1$: the simplest possible Lie algebra produces 
the prototype of all Toda models---the
Liouville theory. This consists of a single bosonic field 
$\phi$ with lagrangian
$$
L_{\exp} (\phi) = 
\del_+ \phi \, \del_- \phi \, - \, \exp 2 \phi \ .
\nfr{bliou}
In the superspace approach, this actually comes accompanied by a free
fermion $\psi$ so that
$$L_{\rm bos} = L_{\exp} (\phi) \ , \quad
L_{\rm ferm} = L_{\rm free} (\psi) \ .
\nfr{sliou}
As for any bosonic algebra, however, the fermionic degrees of freedom 
decouple completely.
\vskip 2pt
\noindent
{\bf (2)} $\b(0,1)$: the simplest possible superalgebra with a single
fermionic simple root.
On choosing a suitable normalization for the fields,
the expressions for $L_{\rm bos}$ and $L_{\rm ferm}$ given in \sliou \
above are unchanged.
But the fermion is no longer free; instead it is coupled to 
the Liouville field through the 
interaction term
$$
L_{\rm int} = \psib \psi \exp \phi \ . 
\nfr{slint}
This is the supersymmetric Liouville theory (see eg.~[\Ref{LM1}] and
references given there). 
\vskip 2pt

\noindent
{\bf (3)} B(0,2): $\R$ has $\tau = \{1\}$ and Cartan matrix
$$ a_{ij} = \pmatrix{1 & -1 \cr -1 & 2 \cr } \ . $$
The full set of equations of motion is
$$
\eqalign{
&\del_+ \del_- \phi_1 = - \exp{2 \phi_1} + \exp{\phi_2} 
+ \Half \psib_{1} \psi_{1} \exp {\phi_1} \ ,
\cr
&\del_+ \del_- \phi_2 = \exp{2\phi_1} - 2 \exp{\phi_2} 
- \Half \psib_{1} \psi_{1} \exp{\phi_1} \ ,
\cr
}
\qquad
\eqalign{
&\del_\pm \psi_{1\mp} = \pm \psi_{1 \pm} \exp{\phi_1} \ , \cr 
&\del_\pm \psi_{2\mp} = \mp \psi_{1 \pm} \exp{\phi_1} \ . \cr
}$$
We can see directly from these equations that 
the bosonic sector is a $\c_2$ 
Toda theory with fields $2\phi_1$ and 
$\phi_2$, confirming the information given in Table 3.
It is also evident that the only relevant fermionic 
field is $\psi_1$, since the combination
$\psi_1 + \psi_2$ decouples, in keeping with Proposition 3. 
Following on from example (2),
this is the next member of a series of conformally-invariant models 
based on the root systems for $\b(0,n)$ [\Ref{WI}]---see Table 2 (a).
For $n > 1$ the models are not supersymmetric; they consist of a $\c_n$
Toda theory, with the field $2\phi_1$ corresponding to the long root,
and $\phi_1$ 
coupled to a single fermion $\psi_1$ through an interaction of type 
\slint .
\vskip 2pt

\noindent
{\bf (4)} A(1,0) = C(2): $\R$ has $\tau = \{1,2\}$ and Cartan matrix
$$
a_{ij} = \pmatrix{ 0 & 1 \cr 1 & 0} \ . $$
The equations of motion now read
$$\eqalign{
\del_+\del_-\phi_1 & = - \exp(\phi_1 + \phi_2) + \Half \psib_2 \psi_2
\exp\phi_2 \ ,
\cr
\del_+\del_-\phi_2 & = - \exp(\phi_1 + \phi_2) + \Half \psib_1 \psi_1
\exp\phi_1 \ ,
\cr}
\qquad
\eqalign{
\del_\pm \psi_{1\mp} & = \pm \psi_{2\pm} \exp{\phi_2} \ , \cr 
\del_\pm \psi_{2\mp} & = \pm \psi_{1 \pm} \exp{\phi_1} \ . \cr
}$$
Since the root system is entirely fermionic, 
this model is supersymmetric. 
The bosonic sector consists of an $\a_1$ Toda, or Liouville theory, 
together with a scalar field. More precisely, if we introduce the fields 
$$
2u = \phi_1 + \phi_2 \quad {\rm and} \quad
2v = \phi_1 - \phi_2 
\nfr{uv} 
then the bosonic sector is 
$$
L_{\rm bos} = L_+ + L_0 = L_{\exp}( u )  - L_{\rm free} ( v ) \ . 
\nfr{abos}
Similarly, if we adopt corresponding 
redefinitions for the fermions
$$
2\chi = \psi_1 + \psi_2 \quad {\rm and} \quad
2 \eta = \psi_1 - \psi_2 
\nfr{chieta}
then we have
$$
L_{\rm ferm} = L_{\rm free} (\chi) - L_{\rm free} (\eta) \ .
\nfr{aferm}
We can complete the description of the model in terms of these new
fields by giving the interaction term 
$$\eqalign{
L_{\rm int} 
& = (\chib \chi + \bar \eta \eta) \exp{u} \cosh v
+ 2 \chib \eta \exp{u } \sinh v \ .
\cr
}\nfr{aint}
This example is known to be closely related to the $N=2$ super Liouville
theory; we will return to it briefly in section 7.
\vskip 2pt

\noindent
{\bf (5)} A(1,0)=C(2): $\R$ has $\tau = \{ 2 \}$ and Cartan matrix 
$$a_{ij} = \pmatrix{ 2 & -1 \cr -1 & 0 \cr } \ . $$
The equations of motion are
$$
\eqalign{
\del_+ \del_- \phi_1 & = - 2 \exp{\phi_1} - \Half \psib_2 \psi_2 \exp
\phi_2 \ , \cr
\del_+ \del_- \phi_2 & = \exp{\phi_1} \ , \cr
} \qquad
\eqalign{
\del_\pm \psi_{1\mp} & = \mp \psi_{2\pm} \exp{\phi_2} \ , \cr 
\del_\pm \psi_{2\mp} & = 0 \ . \cr
}$$
The bosonic sector is again based on $\a_1 \oplus \u(1)$ where now 
$\phi_1$ is the Liouville field and $\phi_1 + 2 \phi_2$ is the scalar,
but apart from this the equations of motion make clear that 
this model bares little resemblance to the previous example based on
the same algebra, eg.~it is not supersymmetric.
This illustrates the fact that the properties of the
Toda theory depend fundamentally on the root system rather than
on the algebra which it defines.
In general there is no reason to expect even that the bosonic sectors
should coincide. Notice also that neither of the fermions are
decoupled in this example, despite the fact that $\psi_2$ obeys a free
equation of motion.
\vskip 2pt

\noindent{\bf (6)} C(3): $\R$ has $\tau = \{1, 2\}$ and Cartan matrix 
$$
a_{ij} = \pmatrix{0&2&-1\cr 2&0&-1\cr -1&-1&2} \ .
$$
The bosonic sector corresponds to $\c_2 \oplus \u(1)$. 
As in example (4), it is useful to introduce the combinations
of fields \uv \ and \chieta .
The expression for the bosonic lagrangian is then similar to \abos \ 
except that $L_+$ is now a lagrangian for a $\c_2$ Toda model with 
fields $u$ and $\phi_3$ while 
$v$ is still the free scalar which
contributes with opposite sign.
Proposition 3 states that $\psi_3$ decouples,
leaving $\chi$ and $\eta$ as the only important fermionic degrees of
freedom, and \aferm \ is unchanged.
The interaction term \aint \ is also unchanged. 
This analysis can be generalized immediately 
to any of the algebras C($m$) with root systems
of type $\dl$. The lagrangian 
$L_+$ becomes a $\c_{m-1}$ Toda model in general, but there are
only ever two relevant fermions which interact with 
the bosons in just the way we have described.
\vskip 5pt

%5
\chapter{Examples without conformal symmetry}

For bosonic Toda theories there is a clear-cut distinction between 
those based on finite-dimensional algebras, which are conformally
invariant, and those based on affine algebras, which are massive.
In the last section we saw that Toda theories based on finite-dimensional
superalgebras are also 
conformally invariant (with the exception of the $\a(m,m)$ series) 
just as in the bosonic case. 
When we come to Toda models based on affine superalgebras, however,
we find a whole range of new possibilities.
Some models resemble their bosonic siblings in that they 
consist entirely of massive degrees of freedom. But others 
contain mixtures of massless and massive modes, and some may
have no massive degrees of freedom at all in their bosonic sectors.
\vskip 15pt 

%%%%%%%%%%%%%%% TABLE4
\noindent{\bf Table 4: Some low-rank linearly-dependent simple root systems}
\vskip 8pt
\settabs\+NoXXX&NameXXXXXX&RXXXXXXXXXXX&B+XXXXXXXXXX&B-XXXXXXX&V0\cr
\+\phantom{XXX}{\bf Superalgebra} && \phantom{XX} 
$\R$ & \phantom{X} $\B_+$ & $\B_-$ & \kern -8pt dim$V_0$ \cr
\vskip 8pt
\+{\bf (1)} &$\a_1^{(1)}$ 
&\hbox{\epsfxsize=32pt\epsfbox{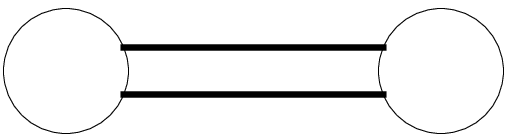}}      
&\hbox{\epsfxsize=32pt\epsfbox{A1a.eps}}      
&\kern 1pt --- 
&0\cr
\vskip 5pt
\+{\bf (2)} &$\a(1,0)^{(2)}$ %\c(2)^{(2)}
&\hbox{\epsfxsize=32pt\epsfbox{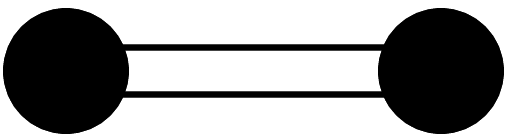}}     
&\hbox{\epsfxsize=32pt\epsfbox{A1a.eps}}      
&\kern 1pt --- 
&0\cr
\vskip 5pt
\+       &$\a(2,0)^{(4)}$ 
&\hbox{\epsfxsize=32pt\epsfbox{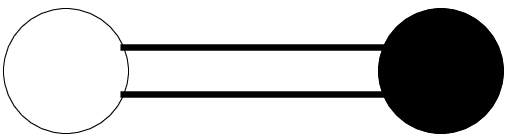}}     
&\hbox{\epsfxsize=32pt\epsfbox{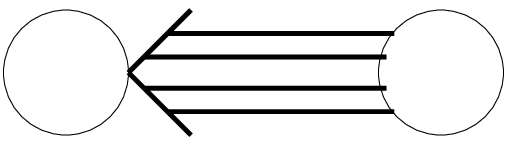}}     
&\kern 1pt ---  
&0\cr 
\vskip 5pt
\+       &$\b(0,1)^{(1)}$ 
&\hbox{\epsfxsize=32pt\epsfbox{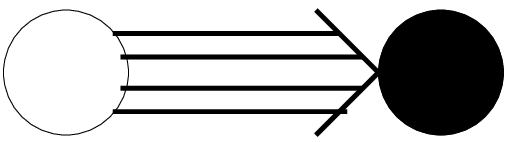}} 
&\hbox{\epsfxsize=32pt\epsfbox{A1a.eps}}  
&\kern 1pt ---  
&0\cr 
\vskip 5pt
\+        &$\a(2,2)^{(4)}$ \ \  
&\hbox{\epsfxsize=56pt\epsfbox{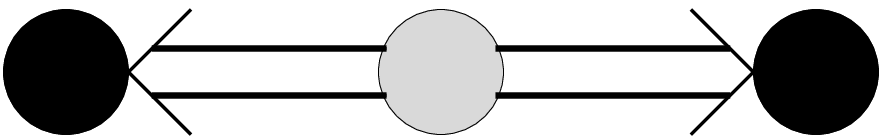}}
&\def\box1{\hbox{\epsfxsize=32pt\epsfbox{A20b.eps}}}
\hbox{\kern 28.5pt \rotf1} 
&\kern -10 pt \hbox{\epsfxsize=32pt\epsfbox{A20b.eps}} 
&0\cr
\vskip 5pt
\+        &$\d(2,1)^{(2)}$ \ \  
&\hbox{\epsfxsize=56pt\epsfbox{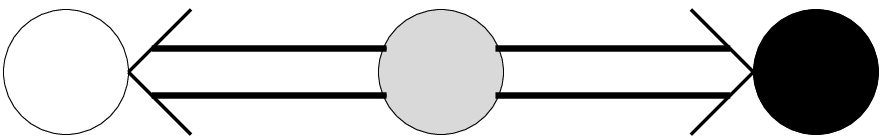}} 
&\hbox{\epsfxsize=32pt\epsfbox{A1a.eps}} 
&\kern 2pt \hbox{\epsfxsize=8pt\epsfbox{p1.eps}} 
&0\cr 
\vskip 5pt
\+        &$\a(2,1)^{(2)}$ \ \ 
&\hbox{\epsfxsize=56pt\epsfbox{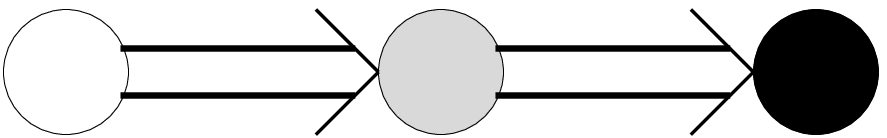}} 
&\def\box1{\hbox{\epsfxsize=32pt\epsfbox{A20b.eps}}}
\hbox{\kern 28.5pt \rotf1} 
&\kern 2pt \hbox{\epsfxsize=8pt\epsfbox{p1.eps}} 
&0\cr
\vskip 5pt
\+{\bf (3)} &$\b(1,1)^{(1)}$ \ \ 
&\lower 12pt \hbox{\epsfxsize=25pt\epsfbox{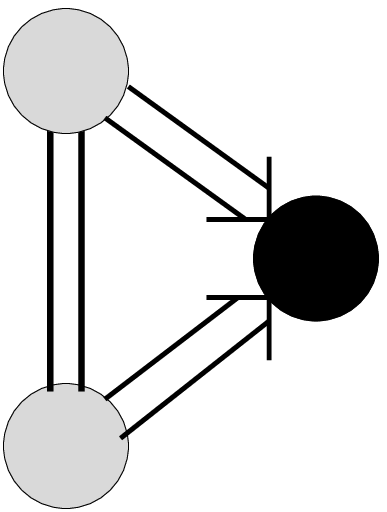}} 
&\hbox{\epsfxsize=32pt\epsfbox{A1a.eps}} 
&\def\box1{\hbox{\epsfxsize=32pt\epsfbox{A1a.eps}}}
\kern -1.3pt \lower 12pt \hbox{\rotl1} 
&0\cr
\vskip 5pt
\+       &$\a(2,1)^{(2)}$  
&\lower 12pt 
\hbox{\epsfxsize=25pt\epsfbox{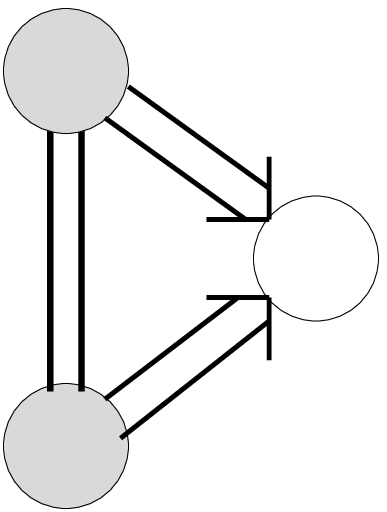}} 
&\def\box1{\hbox{\epsfxsize=32pt\epsfbox{A20b.eps}}}
\hbox{\kern 28.5pt \rotf1} 
&\kern 1pt ---  
&1\cr
\vskip 5pt
\+       &$\a(1,0)^{(1)}$ %c(2)^{(1)} 
&\lower 12pt 
\hbox{\epsfxsize=25pt\epsfbox{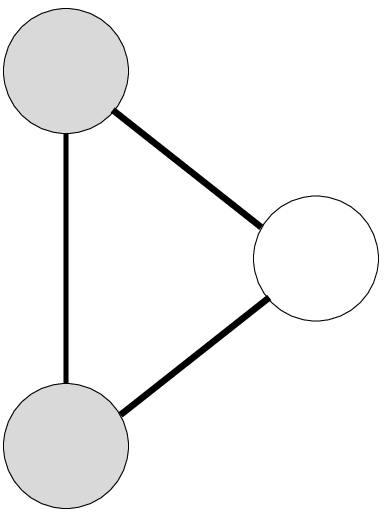}} 
&\hbox{\epsfxsize=32pt\epsfbox{A1a.eps}} 
&\kern 1pt --- 
&1\cr
\vskip 5pt
\+{\bf (4)} &$\a(1,1)^{(1)}$ 
&\lower 12pt 
\hbox{\epsfxsize=32pt\epsfbox{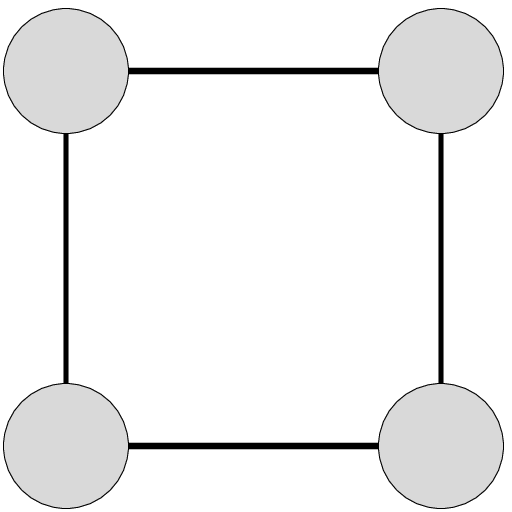}} 
&\hbox{\epsfxsize=32pt\epsfbox{A1a.eps}} 
&\def\box1{\hbox{\epsfxsize=32pt\epsfbox{A1a.eps}}}
\kern -1.3pt \lower 12pt \hbox{\rotl1} 
&0\cr
\vskip 5pt
\+      &$\a(1,1)^{(1)}$  
&\lower 12pt \hbox{\epsfxsize=32pt\epsfbox{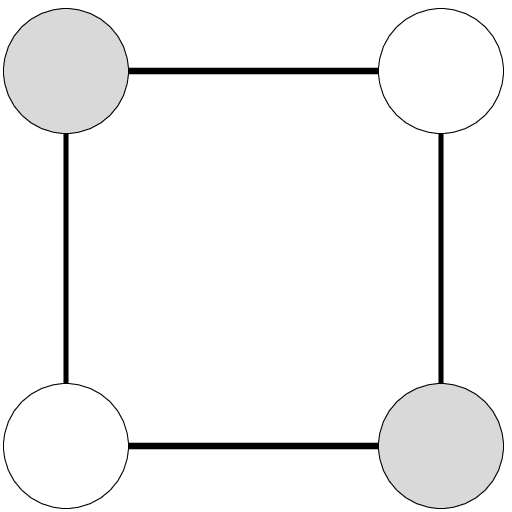}} 
&\kern 11.5pt \hbox{\epsfxsize=8pt\epsfbox{p1.eps}}   
&\kern 2pt \hbox{\epsfxsize=8pt\epsfbox{p1.eps}} 
&0\cr
\vskip 5pt
\+       &$\a(2,0)^{(1)}$ 
&\lower 12pt \hbox{\epsfxsize=32pt\epsfbox{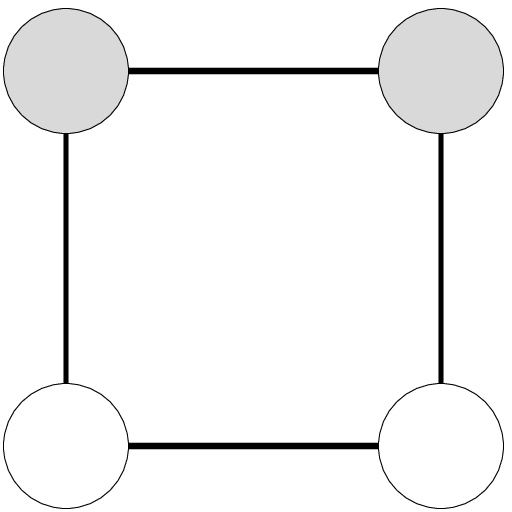}} 
&\lower 12pt \hbox{\epsfxsize=32pt\epsfbox{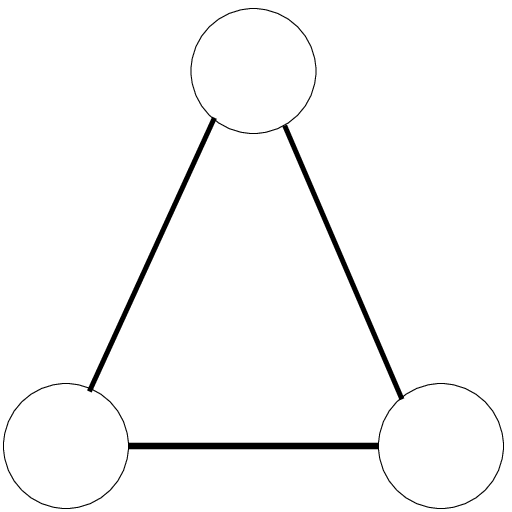}} 
&\kern 1pt --- 
&1\cr
\vskip 5pt
\+        &$\a(3,1)^{(2)}$ 
&\lower 12pt \hbox{\epsfxsize=32pt\epsfbox{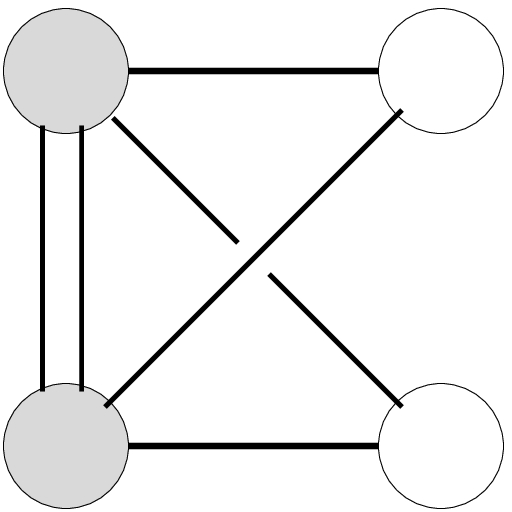}} 
&\kern -10pt \hbox{\epsfxsize=56pt\epsfbox{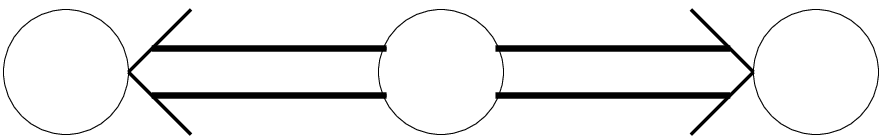}}   
&\kern 1pt ---  
&1\cr
\vskip 5pt
\+{\bf (5)} &$\d(2,1)^{(1)}$  
&\lower 12pt \hbox{\epsfxsize=32pt\epsfbox{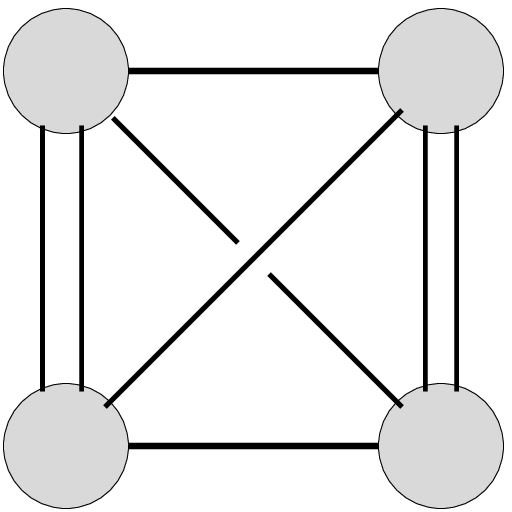}} 
&\hbox{\epsfxsize=32pt\epsfbox{A1a.eps}}   
&\def\box1{\hbox{\epsfxsize=32pt\epsfbox{A1a.eps}}}
\lower 12pt \hbox{\kern -13pt \rotl1 \kern 20pt \rotl1}
&0\cr
\vskip 20pt

We can readily see how these various possibilities arise by applying our 
diagrammatic method to determine the bosonic
sectors of a number of representative examples,
with results summarized in Table 4.
It is generally straightforward to combine the diagrams in 
Figures 2, 3 and 4 in order to extract the necessary information.
The only cases in which a little extra thought may be
required are those root systems 
with a very small number of nodes so that the
tails of the affine diagrams are not well-separated; we have included the
least obvious examples in the table.

To amplify our previous remarks, we note that 
some of these models are purely massive, including those labeled (1)-(5);
some contain mixtures of massive and massless degrees of freedom
(the most trivial possible conformal theory being a free massless
field); and one theory, based on the second choice of simple roots
for $\a(1,1)^{(1)}$,
has a conformally-invariant bosonic sector. Many more examples of each
kind can be found by looking at higher-rank diagrams.
The cases (1)-(5) turn out to be of special interest, so we
now consider them in more detail.

We mentioned in the last section that for $\R$ linearly independent,
$r=n$, we can choose the values of 
$|\mu_i|$ as we please by making constant shifts in the fields. 
In this section we are taking $\R$ to be linearly dependent.
All but one family of simple roots of this type 
have $n=r{+}1$, in common with bosonic affine algebras.
This means that they obey exactly one linear relation
and so, while we may be free to alter {\it ratios\/} of the parameters 
$\mu_i$, there is always one surviving overall mass scale $m$ 
which cannot be eliminated.
The exceptions, with $n=r{+}2$,
are simple root systems for the algebras $\a(m,m)^{(1)}$.
In these cases there are two 
independent mass parameters $m$ and $m'$, 
or equivalently a single mass parameter
$m$ and an additional dimensionless ratio $\gamma = m/m'$ (in
addition to the possibility of introducing the usual kind of Toda 
coupling $\beta$ which we are omitting throughout). 
In all the examples below $\mu_i$ will be chosen
to be a zero-eigenvector of the Cartan matrix, ensuring that the 
minimum of the bosonic potential occurs where the fields vanish.
As with the conformal examples, we adopt overall normalizations so as to
fix various numerical constants in a simple way. 
\vskip 2pt

\noindent
{\bf (1)} $\a_1^{(1)}$: the simplest affine Lie algebra,
with Cartan matrix 
$$a_{ij} = \pmatrix{2&-2\cr-2&2\cr} \ . $$
Note that the rows are proportional to one another, 
implying a linear relation between
the superfields $\Phi_1 + \Phi_2 = 0$. 
This results is the prototype of all affine Toda theories---the
sinh-Gordon theory. There is a single independent 
bosonic field $\phi$ of mass $m$ governed by the lagrangian
$$\eqalign{
L_{\sinh} (\phi;m) & = \del_+ \phi \, \del_- \phi \, - \, m^2 
\sinh^2 \! \phi \cr
&= \del_+ \phi \, \del_- \phi \, - \, 
\Half m^2 (\cosh 2 \phi -1 ) \, ,
\cr}
\efr
where we have added a constant to make the minimum value 
of the potential zero.
As with the Liouville theory, this is accompanied in the superspace
construction by a free
massless fermion
$\psi$ so that
$$L_{\rm bos} = L_{\sinh} (\phi;m) \ , \qquad 
L_{\rm ferm} = L_{\rm free} (\psi) \ , 
\nfr{bsg} 
but the boson and fermion do not interact with one another. 
The relationship with the sine-Gordon theory will be discussed in
section 7. 
\vskip 2pt
\noindent
{\bf (2)} $\a(1,0)^{(2)} = \c(2)^{(2)}$: defined by the same Cartan
matrix as above, 
but now the roots are fermionic and the theory is supersymmetric. 
With a suitable normalization for the fields, 
the terms $L_{\rm bos}$ and $L_{\rm ferm}$ are unchanged, but there is an 
additional interaction term
$$
L_{\rm int} = m \psib \psi \cosh \phi 
\nfr{ssg}
which gives the fermion a mass $m$ equal to that of the boson.
This is the supersymmetric sinh-Gordon theory.
\vskip 2pt

\noindent
{\bf (3)} $\b(1,1)^{(1)}$: $\R$ has $\tau = \{1, 2, 3\}$ and 
Cartan matrix 
$$
a_{ij} = \pmatrix{1&-1&-1\cr -1&0&2\cr -1&2&0\cr} \ .
$$
There is a single linear relation amongst the simple roots,
and consequently the superfields obey $2 \Phi_1 + \Phi_2 + \Phi_3 =
0$. The bosonic sector consists of a pair of 
$\a^{(1)}_1$ Toda models, or sinh-Gordon theories, 
which contribute with opposite signs.
To see this explicitly we 
introduce the independent fields
$$u = - \phi_1 = \phi_1 + \phi_2 + \phi_3 \ , \qquad
2v = \phi_1 + \phi_2 = - (\phi_1 + \phi_3) \ ,
$$
and in terms of these we find 
$$
L_{\rm bos} = L_+ + L_-  = L_{\sinh} (u ; 2m) - 4 L_{\sinh} ( v ; m) \ .
\efr
Note that the masses of these two theories are 
different and also that  
the relative coefficient between their lagrangians cannot be changed.
If we introduce the corresponding fermionic fields 
$$\chi = - \psi_1 = \psi_1 + \psi_2 + \psi_3 \ , \qquad
2 \eta = \psi_1 + \psi_2 = - (\psi_1 + \psi_3) \ ,
$$
then we find similarly 
$$
L_{\rm ferm} = L_{\rm free} (\chi) - 4 L_{\rm free} (\eta) \ .
\efr
The interaction term is also easily found to be
$$
L_{\rm int} =  
m \left \{ \,  \chib \chi \exp(- u) \, + \, 
( \, \chib \chi + 4 \bar \eta \eta \, ) \exp{ u } \cosh 2v  \, + \, 
4 \chib \eta \exp{u } \sinh 2v \, \right \}
\ .
\nfr{bmass} 
\vskip 2pt

\noindent
{\bf (4)} $\a(1,1)^{(1)}$: $\R$ is defined by $\tau = \{1, 2, 3, 4\}$ and
the Cartan matrix 
$$
a_{ij} = \pmatrix{0&1&0&-1\cr 1&0&-1&0\cr 0&-1&0&1\cr -1&0&1&0\cr} \ .
$$
As mentioned earlier, there are two linear relations amongst 
simple roots for these algebras, and in this case we have 
$\Phi_1 + \Phi_3 = \Phi_2 + \Phi_4 = 0$.
Once again the bosonic sector consists of a pair of sinh-Gordon
theories which 
contribute with opposite overall signs.
We introduce 
$$
\eqalign{
2 u  & = \phi_1 + \phi_2 = - (\phi_3 + \phi_4) \ , \cr
2 v  & = \phi_1 + \phi_4 = - (\phi_2 + \phi_3) \ , \cr
}
\qquad
\eqalign{
2 \chi & = \psi_1 + \psi_2 = - (\psi_3 + \psi_4) \ , \cr
2 \eta & = \psi_1 + \psi_4 = - (\psi_2 + \psi_3) \ , \cr
}
$$
and then the bosonic sector is given by
$$
L_{\rm bos} = L_+ + L_- = L_{\sinh} ( u  ; m) - L_{\sinh} ( v ; m) \ ,
\efr
while in the fermionic sector 
$$
L_{\rm ferm} = L_{\rm free} (\chi) - L_{\rm free} (\eta) \ .
\efr
Lastly,  we find the interaction term 
$$\eqalign{
L_{\rm int} 
& = \Half  m (\, \chib \chi + \etab \eta \, ) 
 \left \{ \, \gamma \cosh ( u + v) + \gamma^{-1} \cosh (u  - v ) \,
\right \} \cr
& \qquad 
+ m \chib \eta \left \{  \gamma \cosh (u + v ) - \gamma^{-1} 
\cosh(u  - v ) \, \right \} \cr
& = m (\chib \chi  + \etab \eta ) \cosh u \cosh v \, + \, 
2 m \chib \eta \sinh u \sinh v \quad {\rm when} \ \gamma =
1 \ . \cr
}\nfr{amass} 
Notice that the masses of the sinh-Gordon theories appearing in
$L_\pm$ are equal, 
irrespective of the value of the dimensionless parameter $\gamma$.  
The case $\gamma = 1 $ is given explicitly in preparation for 
some observations to follow in Section 7.
\vskip 2pt

\noindent
{\bf (5)} $\d(2,1)^{(1)}$: $\R$ defined by 
$\tau = \{1, 2, 3, 4\}$ and Cartan matrix
$$
a_{ij} = \pmatrix{0&2&-1&-1\cr2&0&-1&-1\cr -1&-1&0&2\cr -1&-1&2&0\cr}
\ .
$$
Here there is one linear relation 
$\Phi_1 + \Phi_2 + \Phi_3 + \Phi_4 = 0$.
In this case the bosonic sector consists of three sinh-Gordon
theories, one contributing positively, the others negatively.
We introduce the independent fields
$$
\eqalign{
2 u \,&= \phi_1 + \phi_2 = - (\phi_3 + \phi_4) \ , \cr
2 v_1 &= \phi_1 + \phi_3 = - (\phi_2 + \phi_4) \ , \cr
2 v_2 &= \phi_1 + \phi_4 = - (\phi_2 + \phi_3) \ , \cr
}
\qquad
\eqalign{
2 \chi \, &= \psi_1 + \psi_2 = - (\psi_3 + \psi_4) \ , \cr
2 \eta_1 &= \psi_1 + \psi_3 = - (\psi_2 + \psi_4)  \ , \cr
2 \eta_2 &= \psi_1 + \psi_4 = - (\psi_2 + \psi_3)  \ . \cr
}
$$
Then the bosonic sector is
$$
L_{\rm bos} = L_+ + L_-  
= L_{\sinh} (u ; 2m) - 2 L_{\sinh} (v_1 ; m) - 2 L_{\sinh} (v_2 ; m) \
,
\nfr{dbos}
and in the fermionic sector
$$
L_{\rm ferm} = L_{\rm free} (\psi) - 2 L_{\rm free} (\eta_1) - 2
L_{\rm free} ( \eta_2 ) \ .
\nfr{dferm}
A short calculation reveals the interaction term 
$$\eqalign{
L_{\rm int} 
%& = m \psib_{1} \psi_{1} \exp{\phi_1} +  
%m \psib_{2} \psi_{2} \exp{\phi_2} + m \psib_{3} \psi_{3} \exp {\phi_3} 
%+ m \psib_{4} \psi_{4} \exp {\phi_4} \cr
& = m (\chib \chi + \etab_{1} \eta_{1} + \etab_{2} \eta_{2} ) 
\{ \, \exp u \cosh (v_1 + v_2 )
+ \exp {(-u)} \cosh (v_1 - v_2 ) \, \} \cr
&\qquad \qquad + 2m \etab_1 \eta_2  
\{ \, \exp u \cosh (v_1 + v_2)
- \exp {(-u)} \cosh (v_1 - v_2 ) \, \} \cr
&\qquad \qquad + 2m \chib  \eta_{1} 
\{ \, \exp u \sinh (v_1 + v_2 )
- \exp {(-u)} \sinh (v_1 - v_2) \, \} \cr
&\qquad \qquad + 2m \chib \eta_{2}  
\{ \, \exp u \sinh (v_1 + v_2)
+ \exp {(-u)} \sinh (v_1 - v_2 ) \, \}  \ . \cr
}\nfr{dmass}
If we set $v_1=v_2=v$ and $\eta_1 = \eta_2 =\eta$ then
we recover example (3) above. This is because the simple root systems 
involved are related by the folding operation discussed for
superalgebras in [\Ref{FSS}].
\vskip 5pt

Given everything  we have said so far, it is natural to ask 
which Toda theories based on superalgebras are purely massive. 
\vfill \eject 
\noindent
{\bf Proposition 6:} A Toda theory is purely massive iff it is based on 
a simple root system $\R$ whose Dynkin diagram belongs to one of 
the following classes: 
\hfil \break
{\bf (a)} Affine diagrams with no grey nodes---in
addition to the usual bosonic affine root systems we have 
precisely the cases listed in Table 2, section (b).
\hfil \break
{\bf (b)} Affine diagrams which are entirely
fermionic, ie.~which give massive supersymmetric theories---precisely the
cases listed in Table 1, section (c).
\vskip 2pt
\noindent
Proof: In order to get a
purely massive bosonic theory there must be a linear relation amongst the
roots in both $\B_+$ and $\B_-$, so that they both represent affine bosonic 
algebras; this means the numbers of nodes in these
diagrams, denoted by $n$, 
must exceed the dimensions of $V_\pm = \langle \B_\pm
\rangle$. But in addition $\B_+$ and $\B_-$ must 
together span $V$ to ensure that 
$V_0$ is trivial and so exclude massless scalars. 
Let us consider first those diagrams with $n = r+1$,
which omits just the root systems of $\a(m,m)^{(1)}$. 
We have one linear relation on $\R$ and hence at least one on $\B_+ \cup \B_-$.
If either of $\B_\pm$ is empty, then this single
linear relation can suffice to make the remaining bosonic root system 
affine---this gives rise to the class of allowed cases in (a).
If neither of $\B_{\pm} $ is trivial, however, then we need one
extra linear relation to ensure that both
represent bosonic affine root systems.
For all allowed Dynkin diagrams, it is easy to see that the number of 
bonds never exceeds the number of nodes by more than one.   
Using this in conjunction with the definitions of $\B_\pm$, 
we see that $n(\B_+) + n(\B_-) \leq n(\R) + 1$, with equality only
when $\R = \F$ is purely fermionic.
This gives rise to almost all the examples in class (b).
Finally, we must consider root systems 
for $\a(m,m)^{(1)}$ with $n= r + 2$. But these diagrams are all circular, 
which means that the bound used above can be
improved to $n(\B_+) + n(\B_-) \leq n(\R) $, again with equality only
for the purely fermionic case. This allows just the last entry in
Table 2, completing the proof.
\vskip 5pt

The examples (1)-(5) from Table 4 
all belong to one or other of the classes in Proposition 6.
We should emphasize that for class (b)---supersymmetric
models---the lagrangians are generally of indefinite signature.
Nevertheless, from what we have said 
in Section 3, the masses of the bosons are perfectly well-defined,
since the relative sign between the potential and kinetic 
energy terms is always physically sensible (provided we choose the
parameters $\mu_i$ as indicated in Proposition 2). 
Furthermore, the lists of bosonic sectors given in Tables 1 and 2 tell us 
immediately the bosonic masses of any of 
the superalgbera theories in Proposition 6,
in terms of the known results for bosonic Toda models 
[\Ref{MOP},\Ref{BTS},\Ref{FLO}]. 
This provides a proof of the 
mass formulas for the particular series 
$\a(m,m)^{(1)}$, $\b(m,m)^{(1)}$, and $\d(m{+}1,m)^{(1)}$ 
which were conjectured in [\Ref{QI}] on the basis of computer calculations. 

For the supersymmetric examples in class (b) of Proposition 6, 
the fermions must be degenerate in mass with the bosons.
Whether these models can be associated with 
`pseudo-unitary' quantum 
S-matrices (see eg.~[\Ref{CM},\Ref{TH}]) as suggested by perturbative
calculations [\Ref{QI}] is a fascinating issue, but not one that we shall
attempt to pursue here.
By contrast, the examples in class (a) have
conventional positive-definite lagrangians and so they 
can be quantized straightforwardly. Since they are not supersymmetric,
the bosonic and fermionic masses will be different.
Some relevant S-matrices are discussed in [\Ref{STS}].
\vfill \eject

%6
\chapter{Conformal curiosities}

We stated in Proposition 5 that the existence of a conformal
symmetry \conf \ relies on $\R$ being linearly independent.
But there are 
numerous Toda models for which $\R$ is linearly dependent 
and yet for which the bosonic sector $L_{\rm bos}$ is conformally
invariant when taken in isolation.
Perhaps the most striking examples are those based on 
the finite-dimensional algebras
$\a(m,m)$ whose special properties were noted in section 4,
though we have seen similar behaviour 
for several other models based on infinite-dimensional superalgebras
in the last section.
In all these cases we must conclude that the interactions with the
fermions break the conformal invariance present in the bosonic sector.
It is interesting to consider whether the conformal symmetry
might be extended in some more subtle way so as to include the fermions.

Keeping to our earlier notation, we write $\phi_j = \alpha_j \cdot
\phi$ and $\psi_j = \alpha_j \cdot \psi$, and 
we consider the transformations 
$$
\phi_{j} \rightarrow \phi_{ j } + h_j \log (\del_+ y^+
\del_- y^-)\, , \qquad 
\psi_{j \pm} \rightarrow (\del_\pm y^\pm)^{1-h_j} \psi_{j \pm} \ ,  
\nfr{newconf}
where $h_j$ may be a different real number for each 
fermionic root $\alpha_j$. 
The usual choice, corresponding to \conf , is $h_j = 1/2$. 
It is easy to see from \clag \ that 
the relationship between the 
conformal weights of the bosons and fermions in \newconf \
automatically renders the 
terms in $L_{\rm int}$ invariant.
The problem is that it is much more difficult to achieve 
invariance of the other terms---particularly since a 
linear relation amongst the simple roots imposes a corresponding
relationship amongst the fields and hence amongst the possible values of the 
numbers $h_j$.
There are, however, two families of examples in which exotic
conformal symmetries of this type exist.

First we consider a Toda model based on $\a(m,m)$, 
$\R$ being of type $\au$ with $2m{-}1$ nodes.
Examination of the Cartan matrix reveals that 
the linear relation involves only the odd-numbered roots:
$\sum_{j} \alpha_{2j+1} = 0$. 
Now it is easy to see that in this case \clag \ is invariant if we 
choose $h_j = 0$ for $j$ odd and $h_j = 1$ for $j$ even,
and that this is consistent with the linear relation amongst the
fields. (Here `odd' and `even' refer to the 
values of the integer labels, not to the grading of the roots, which
are all fermionic!)
This was pointed out in [\Ref{EH2}] for the simplest case
based on A(1,1).
The second set of examples is based on $\a(m,m{-}1)$,
again with $\R$ of type $\au$ but this time with $2m$ nodes.
Now the simple roots are actually linearly independent,
and we have already seen that this model is invariant under the
standard conformal transformations with $h_j = 1/2$. 
But in fact it is invariant under {\it any\/} assignment of 
conformal weights of the type
$$
h_j = s \ {\rm for} \  j \ {\rm odd} \ , 
\qquad h_j = 1{-}s \ {\rm for} \ j \ {\rm even} \ ,
\efr
where $s$ is any real number.

The existence of these rather strange conformal symmetries can be
traced to a specific feature of the 
simple root systems of type $\au$. 
They can be divided into disjoint sets 
$\F_0 = \{ \alpha_j \in \F \ | \ j \ {\rm even} \ \}$ 
and $\F_1 = \{ \alpha_j \in \F \ | \ j \ {\rm odd} \ \}$ and on 
each of these individually the inner-product vanishes, so we obtain 
a non-zero result only when we choose one vector from each set.
For this to hold it is obviously necessary 
that all the simple roots are null, but this is by no means
sufficient, and it is not difficult to see that the 
$\au$ root systems are the unique ones having this property.
The reason it is important is that
the conformal weights of a fermion are usually fixed to be 1/2 
by its kinetic terms, but when we have inner-products of this 
peculiar type a fermion field 
never appears paired with itself in the kinetic term, because of
the degeneracy. 

These remarks strongly suggest that 
non-standard conformal symmetries \newconf \ 
with $h_j \neq 1/2$ occur only for the models that we have just discussed
based on the $\au$ root systems. 
Although there are many other Toda models based on linearly-dependent
simple root systems with conformally-invariant 
bosonic sectors,
it seems likely that for these the conformal symmetry is irrevocably 
broken by the addition of fermions.
This is similar in many ways to some recent results concerning the
incompatibility of (1,0) supersymmetry and extended conformal invariance in
Toda models [\Ref{EM2}]. 

The emergence of fermions with integer spin is reminiscent of 
ghosts in a topological field theory.
Indeed, the special feature of the $\au$ root system---that it can be 
divided into the two disjoint degenerate sets $\F_0$ and $\F_1$---is
also known to be linked to the existence 
of $N=2$ supersymmetry in certain versions of the $\a(m,m{-}1)$
models [\Ref{EH1},\Ref{JE},\Ref{EH2}]. 
It is natural to expect that the family of conformal symmetries 
which exists between 
$s=1/2$ and $s=0$ or 1 should correspond to the usual 
twisting of an $N=2$ model into a topological theory.
The fact that the conformal symmetry of the $\a(m,m)$ model exists
only for $s=0$ means that it does not seem to be immediately related
to a topological twisting; nevertheless it would be interesting to
investigate whether it too
might have a topological character.
Topological properties of massive Toda theories have already been
investigated in [\Ref{PPZ}].

\chapter{Twisted reality conditions}

We now discuss one last ingredient which is important in 
the taxonomy of Toda models.
We have assumed so far that the fields appearing in the Toda equations
are real. 
In fact the integrability of the equations is unaffected if we take all
the fields to be complex in \eqm , so there exist complex versions of
all the Toda theories we have been considering. 
A related possibility which is more interesting in many respects is 
to impose non-standard or 
{\it twisted\/} reality conditions
on the fields, corresponding to a reflection symmetry of the
simple root system $\R$.
This was first pointed out for abelian Toda models in [\Ref{JE}]
and applied to the study of $N=2$ supersymmetry in [\Ref{JE},\Ref{EH2}];
the construction has recently been related to Hamiltonian reduction
and extended to the case of non-abelian Toda models in [\Ref{EM1}].

In the bosonic case, a symmetry of $\R$ is equivalent to a symmetry of the
Dynkin diagram. For superalgebras, 
however, we encounter
once again the problem that the diagram may have more 
symmetry than the Cartan matrix, because of the now familiar issue of
signs. 
With the approach we have developed in this paper, we 
can extend the results of [\Ref{JE}], clarifying the effect on the
bosonic sector. 
\vskip5pt
\noindent 
{\bf Proposition 7:} Let $\sigma$ be a permutation of order two
acting on a set of simple roots $\R$.
\hfil \break
{\bf (a)} We can impose 
reality conditions 
$$\alpha_i \cdot \Phi^* =
\alpha_{\sigma(i)} \cdot \Phi\nfr{twist}
consistent with integrability of the
Toda equations based on $\R$ provided: 
$$a_{\sigma(i) \sigma(j)} = a_{ij} \quad {\rm and} \quad
\mu_{\sigma(i)} = \mu_i \ . \nfr{twsym}
The first condition means that $\sigma$ 
can be regarded as a reflection symmetry of the Dynkin diagram 
{\it even after attaching signs\/} as described in 
Section 2. The second condition is 
automatically satisfied if we have chosen $|\mu_i| = 1$ when $n=r$,
or if we have chosen $\mu_i$ to coincide with the unique null
eigenvector of the Cartan matrix when $n=r{+}1$.
(We have assumed $\mu_i$ real, but there is actually no loss of
generality in doing this.)
\hfil \break
{\bf (b)} The permutation $\sigma$ 
acts in an obvious way as a symmetry of 
the associated bosonic root systems $\B_\pm$.
The bosonic sector of the theory defined by \twist \ therefore 
consists of bosonic Toda theories, as explained in Proposition 2, but
each with twisted reality 
conditions corresponding to these symmetries of $\B_\pm$.
\hfil \break
{\bf (c)} Taking standard reality conditions, $\sigma =1$, or twisted reality
conditions, $\sigma \neq 1$, does not affect whether the theory is
conformally-invariant (under \conf ), or whether it is massive,
in which case the classical mass spectrum is unchanged.
\vskip 2pt

\noindent
Proof: The results all follow easily from \eqm , \cem \ and the definition
of $\B_\pm$. Notice that the conformal transformations in \conf \ 
only change the real parts of the fields, no matter what reality
conditions are chosen.
The mass spectrum is obtained by linearizing the equations of
motion \cem \ and so it is clear that this is unchanged.
\vskip 5pt

The imposition of twisted reality conditions 
amounts to taking certain components of the Toda field to be
imaginary rather than real, but in a way which preserves the reality
of the lagrangian. 
This can be thought of as a kind of generalization of the simple behaviour of 
a free massless scalar field, for which
$$
\phi = i \rho \quad \Rightarrow \quad
L_{\rm free} (\phi) = - L_{\rm free} (\rho ) \ .
\nfr{twb} 
All bosonic Toda theories have lagrangians of definite signature, 
so changing the reality conditions of the fields must always
lead to a lagrangian of indefinite signature,
except perhaps when there is only one independent field, as for the
free scalar above.
The behaviour \twb \ cannot be generalized to the Liouville theory,
because there is no non-trivial symmetry of the Dynkin diagram for $\a_1$,
but it can be generalized to its affine counterpart, the sinh-Gordon theory,
based on the algebra $\a_1^{(1)}$. 

We discussed the sinh-Gordon theory as example (1) from Table 4,
and we saw that the model consists of a single bosonic field 
(since the accompanying free fermion can essentially be ignored) 
with lagrangian
$$\eqalign{
L_{\sinh}(\phi;m) & = \del_+ \phi \, \del_- \phi - m^2 \sinh^2 \! \phi
\cr
& = \del_+ \phi \, \del_- \phi - \Half m^2 (\cosh2 \phi - 1) \ .\cr
}\nfr{shine} 
There is clearly a symmetry $\sigma$ of the Dynkin diagram
which exchanges its two nodes and corresponds to reflection in a
vertical axis through the figure drawn in Table 4. The imposition 
of reality conditions \twist \ corresponding to this symmetry 
means that we must set 
$\phi = i \rho$ where $\rho$ is real. 
But 
$$
\phi = i \rho \quad \Rightarrow \quad
L_{\sinh} (\phi  ; m) = - L_{\rm sin} (\rho ; m) \ ,
\nfr{twsg}
where we introduce the well-known sine-Gordon lagrangian
$$\eqalign{
L_{\rm sin} ( \rho ; m ) & = 
\del_+ \rho \, \del_-  \rho  -  m^2 \sin^2 \! \rho \cr
& = \del_+ \rho \, \del_-  \rho  +  \Half m^2 (\cos 2 \rho -1) \ . \cr
}\nfr{sine}
Twisted reality conditions are therefore important 
in understanding how the sine-Gordon theory fits into the general
framework of the Toda construction. 
We repeat that this is the only bosonic example 
for which a lagrangian of definite signature (albeit negative) emerges 
when twisted reality conditions are used; all other bosonic theories acquire 
indefinite signature.

These observations are easily extended to the supersymmetric
sinh-Gordon theory, example (2) of Table 4. Again the symmetry 
corresponds to reflection in vertical axes through the diagrams in the
Table. 
When \twist \ is applied to the fermions we must take 
$\psi = i \lambda$ with $\lambda$ real.
But then 
$$
\psi = i \lambda \quad \Rightarrow \quad
L_{\rm free} (\psi) = - L_{\rm free} (\lambda) \ .
\nfr{twf} 
(Strictly speaking this applies in the previous example too, although there the
fermions were ultimately irrelevant.)
The interaction term \ssg \ also remains real, becoming 
$$
L_{\rm int} = - m \bar \lambda \lambda \cos \rho
\efr
when
expressed in terms $\rho$ and $\lambda$. (See [\Ref{Ahn}] for the
super sine-Gordon S-matrix.) 

When we consider other superalgebra models, we find that the situation
encountered for bosonic theories may be reversed: starting from a model 
with an indefinite lagrangian, 
it may be possible to choose reality conditions 
which render the lagrangian positive-definite.
The superalgebra models that have positive-definite actions with
conventional reality conditions were studied by Olshanetsky [\Ref{O}];
these are the theories to which we have already drawn attention in 
Table 2.
In addition, it was pointed out in [\Ref{JE}] that the family of
conformal models based on the algebras $\c(m)$ with simple roots of
type $\dl$ have indefinite signature with conventional reality
conditions, but  become positive-definite when twisted reality
conditions are used.

The first two members of this conformal series were discussed  
as examples (4) and (6) in Table 3 (recall that $\a(1,0) = \c (2)$).
Referring to the explicit descriptions of these models and their 
Cartan matrices given there, 
we clearly have a reflection symmetry $\sigma$ 
defined by exchange of the roots labelled in each case by 1 and 2.
The corresponding reality condition \twist \ implies that 
$u$ and $\chi$ should still be real, but now $v=iw$ and $\eta=i\lambda$ 
where $w$ and $\lambda$ are real. 
For the lagrangian written in \abos \ we see that $L_+$ is
unchanged but 
$L_0 = - L_{\rm free} (v) = L_{\rm free} (w) $
so that $L_{\rm bos}$ becomes positive-definite.
Similarly in the fermionic sector \aferm \ we have
$L_{\rm free} (\eta) = - L_{\rm free} (\lambda)$
so that $L_{\rm ferm}$ is now also positive-definite.
Lastly, the interaction term \aint \ becomes 
$$
L_{\rm int} = (\chib \chi - \bar \lambda \lambda) \exp u \cos w
 -2 \chib \lambda \exp u \sin w \ .
\efr
The case C(2) is the $N=2$ super-Liouville theory; for more details, including 
the $N=2$ superspace description, see eg.~[\Ref{EH2},\Ref{LM2}].

In [\Ref{JE}] it was shown that C($m$) is the only family of conformal
theories for which twisted reality conditions lead to
positive-definite actions in the way we have just described.
The techniques we have developed in this paper allow one to prove this
in a much simpler way, and also to extend the 
analysis beyond the super sinh/sine-Gordon theory to all
affine superalgebras.
\vskip 5pt
\noindent
{\bf Proposition 8:} The simple root systems $\R$ which yield
positive-definite Toda lagrangians in conjunction with twisted reality 
conditions are those listed in Table 5.
\vskip 15pt

{\bf Table 5: Root systems with positive-definite twisted reality
conditions }
\vskip 10pt
\settabs\+XX&XXXX&SuperalgebraXXXXXX&Simple &root
systemXXXXX&Bosonic subalgebra\cr

{\bf (a) Conformal theories}
\vskip 2pt
\+ &&{\bf Superalgebra} & {\bf Simple root system} && 
{\bf Bosonic sector} \cr
\vskip 2pt

\+&& $\a(1,0) = \c(2)$  &\ \ $\au$ & \ \ $2$ nodes & $\a_1 \oplus \u(1)$ \cr
\+&& $\c(m{+}1)$    &\ \ $\dl$ & \ \ $m{+}1$ nodes & $\c_m \oplus \u(1)$ \cr
\vskip 10pt

{\bf (b) Massive theories}
\vskip 2pt
\+ &&{\bf Superalgebra} & {\bf Simple root system} && 
{\bf Bosonic sector} \cr
\vskip 2pt

\+&& $\a_1^{(1)}$   & B-B &\ \ $2$ nodes & $\a_1^{(1)} $ \cr
\+&& $\c(2)^{(2)}$ & $\bu$-$\bu$ &\ \ $2$ nodes 
& $\a^{(1)}_1 $ \cr
\+&& $\a(1,1)^{(1)}$  & $\widehat \au$ &\ \ $4$ nodes 
& $\a^{(1)}_1 \oplus \a^{(1)}_1 $ \cr
\+&&$\b(1,1)^{(1)} $  & $\du$-$\bu$ &\ \ $3$ nodes 
& $\a_{1}^{(1)} \oplus \a_1^{(1)}$\cr
\+&&$\d(2,1)^{(1)} $  & $\du$-$\du$ &\ \ $4$ nodes 
& $\a^{(1)}_{1} \oplus \a^{(1)}_{1} \oplus \a^{(1)}_{1}$\cr
\+&&$\d(2,1;\alpha)^{(1)} $  & &--- 
& $\a^{(1)}_{1} \oplus \a^{(1)}_{1} \oplus \a^{(1)}_{1}$\cr
\vskip 10pt

{\bf (c) Mixed conformal and massive bosonic sectors}
\vskip 2pt
\+ &&{\bf Superalgebra} & {\bf Simple root system} && 
{\bf Bosonic sector} \cr
\vskip 2pt

\+&& $\c(m)^{(1)}$ & $\dl$-C &\ \ $m{+}1$ nodes 
& $\c_{m-1}^{(1)} \oplus \u(1) $ \cr
\+&& $\a(2m{-}1,1)^{(2)}$  & $\dl$-D &\ \ $m{+}2$ nodes 
& $\a_{2m-1}^{(2)} \oplus \u(1) $ \cr
\+&&$ \a(2m{-}2,1)^{(2)}$    & $\dl$-B &\ \ $m{+}1$ nodes  
& $\a_{2m-2}^{(2)} \oplus \u(1)$ \cr
\+&&$\b(1,m)^{(1)} $  & $\dl$-$\bl$ &\ \ $m{+}2$ nodes 
& $\c_{m}^{(1)} \oplus \u(1)$ \cr
\+&&$\d(2,m)^{(1)} $  & $\dl$-$\dl$ &\ \ $m{+}3$ nodes 
& $\c_{m}^{(1)}\oplus\u(1)\oplus\u(1) $ \cr
\vskip 10pt

\noindent
For all entries in the table, $m \geq 2$.
The notation used for Dynkin diagrams combines the building 
blocks introduced in Figures 2, 3 and 4. 
\vskip 2pt

\noindent
Proof: 
The theory will have definite signature only if 
one of the systems $\B_{\pm}$ is either empty or 
corresponds to $\a_1^{(1)}$.
On any other kind of bosonic subsystem, the symmetry must act trivially.
Together these facts tightly constrain the possible Dynkin diagrams
which need be considered.
But even when these conditions are met, we must
still check that the scalar fields are also given the correct signs 
by the choice of reality condition. These observations quickly narrow
down the possibilities to those listed in Table 5, which can then be
checked case by case. 
\vskip 5pt 

Notice that the last two classical entries in section (b) of Table 5 are 
limiting cases of the last two infinite families in section (c).
It is important to treat them separately, however, since they behave 
differently from the other members of their families.
This is because there are more connected fermionic nodes
in their Dynkin diagrams, and so their bosonic sectors, as found by our
diagrammatic methods, deviate slightly from the general patterns found
for the infinite series. 
In particular, the scalar fields corresponding to the U(1) factors 
which occur for the infinite
families are absent in these two special cases.
This is in accordance with Proposition 6, which states that 
a theory is purely massive if it is based on an affine system of 
fermionic simple roots. 

The families in section (c) of Table 5 
may prove interesting to investigate in the
future, since they involve a mixture of massive and massless
degrees of freedom. Their structure, including the interaction potentials,
can easily be found explicitly by
adapting what we have already said about the conformal cases based on
C(2), C(3) and the generalization to $\c(m)$. 
Of most immediate interest, however, are the models 
singled out in section (b) of Table 5. The first two cases 
are the sinh/sine-Gordon theory and its supersymmetric extension,
which we discussed above.
In addition, we find the examples (3) (4) and (5) from our list given
earlier in Table 4.
In each of these cases the new reality conditions
correspond to a reflection symmetry
in a horizontal axis through the diagrams drawn in Table 4,
and the orientations of the diagrams for $\B_\pm$ then
reveal how this symmetry acts on these associated sets of bosonic
roots.

In example (4) it is important that $\gamma = 1 $ in order to fulfill
the second criterion in \twsym \ of Proposition 7.
Using the notation introduced earlier, 
the new reality conditions correspond to setting 
$v=iw$ and $\eta = i \lambda$.
The bosonic lagrangian becomes positive-definite because
$L_- = - L_{\sinh} (v,m) = L_{\rm sin}(w,m)$ giving 
$$
L_{\rm bos} = L_{\sinh} (u;m) + L_{\sin} (w;m) 
\efr
so that we end up with independent sinh-Gordon and sine-Gordon theories.
Similarly, in the fermionic sector $L_{\rm free} (\eta) = - L_{\rm
free} (\lambda)$ so that 
$$
L_{\rm ferm} = L_{\rm free} (\eta) + L_{\rm free} (\lambda) 
\efr
is also positive-definite. The interaction term \amass \ becomes
$$
L_{\rm int} = m ( \chib \chi - \bar \lambda \lambda ) \cosh u \cos w -
2 m \chib \lambda \sinh u \sin w 
\efr
which is clearly real, as it should be.
This is the $N=2$ super sine/sinh-Gordon theory (see [\Ref{KU}] 
for a discussion of the S-matrix).

Examples (3) and (5) are related, as we remarked earlier.
Concentrating on the latter case, the new reality conditions 
are $v_j = i w_j$ and $\eta_j = i \lambda_j$ with $j = 1,2$, 
where the new fields are all real.
We then find a bosonic sector consisting of a 
single sinh-Gordon model and two sine-Gordon models:
$$
L_{\rm bos} = L_{\sinh} (u;2m) + 2 L_{\sin} (w_1;m) + 2 L_{\sin}
(w_2;m) \ , 
\efr
while the fermionic lagrangian is also positive-definite: 
$$
L_{\rm ferm} = L_{\rm free } (\chi) + 2 L_{\rm free} (\lambda_1) 
+ 2 L_{\rm free } (\lambda_2) \ , 
\efr 
and the interaction term \dmass \ becomes
$$\eqalign{
L_{\rm int} 
& = m (\chib \chi - \bar \lambda_1 \lambda_1 - \bar \lambda_2 \lambda_2 ) 
\{ \, \exp u \cos (w_1 + w_2 )
+ \exp {(-u)} \cos (w_1 - w_2 ) \, \} \cr
&\qquad \qquad - 2m \bar \lambda_1 \lambda_2  
\{ \, \exp u \cos (w_1 + w_2)
- \exp {(-u)} \cos (w_1 - w_2 ) \, \} \cr
&\qquad \qquad - 2m \chib  \lambda_{1} 
\{ \, \exp u \sin (w_1 + w_2 )
- \exp {(-u)} \sin (w_1 - w_2) \, \} \cr
&\qquad \qquad - 2m \chib \lambda_{2}  
\{ \, \exp u \sin (w_1 + w_2)
+ \exp {(-u)} \sin (w_1 - w_2 ) \, \}  \ . \cr
}\efr
The generalization to $\d(2,1;\alpha)^{(1)}$ will be considered in detail
elsewhere [\Ref{EM3}].
It would be very interesting to find the S-matrices corresponding to these
new models.

\chapter{Concluding Remarks}
In this paper we have attempted to develop a general and coherent
approach to the analysis of Toda systems based on Lie
superalgebras. 
The observation at the heart of the paper concerns the relationship between
the simple root system $\R$ for a superalgebra and the two associated
bosonic root systems $\B_\pm$,
which allows one to read off the bosonic content of the integrable model.
As part of this discussion, we also hope to
have clarified a number of issues regarding Lie superalgebras, simple
root systems, and their Dynkin diagrams, which may be of interest 
quite independent of any connection to integrable systems.

As far as Toda theories themselves are concerned, one obvious problem
is to extend the methods developed here to the non-abelian situation. 
There are also a number of new models we have found 
which would be very interesting to
study further at the quantum level, particularly the supersymmetric 
combinations of sine/sinh-Gordon theories discussed in the last
section.
\vskip 10pt

\noindent
{\bf Acknowledgements}

\noindent
JME is grateful to Prof.~V.~Kac for helpful correspondence concerning Lie
superalgebras, and in particular for directing him to [\Ref{Leur}].
The research of JME is supported by a PPARC Advanced Fellowship.

\vfill \eject

\references

\beginref

%%%%% Lie algebras and superalgebras

\Rref{LA}{J.E. Humphries, {\sl Introduction to Lie Algebras and 
Representation Theory}, \hfill \break 
(Springer-Verlag, 1972);
\newline
J.P. Serre, {\sl Complex Semi-Simple Lie Algebras},
(Springer-Verlag, 1986)}

\Rref{GO}{P. Goddard and D. Olive, Int. J. Mod. Phys. {\bf A1} (1986) 303}

\Rref{Kac0}{V.G. Kac, {\sl Infinite Dimensional Lie Algebras}, 3rd
Edition, (CUP, 1990)}

\Rref{Kac1}{V.G. Kac, Adv. Math. {\bf26} (1978) 8}

\Rref{Kac2}{V.G. Kac, Adv. Math. {\bf30} (1978) 85}

\Rref{Kac3}{V.G. Kac, {\sl Contravariant form for infinite-dimensional
Lie algebras and superalgebras}, in {\sl Group Theoretical Methods in
Physics XIII}, 
Springer Lecture Notes in Physics, Vol.~94 (1979) 441}

\Rref{Serg}{V.V. Serganova, Math. USSR Iz. {\bf24} (1985) 539}

\Rref{Leur}{J.W. van de Leur, Commun. Alg. {\bf 17(8)} (1989) 1815;
\newline
{\sl Contragredient Lie Superalgebras of Finite Growth},
Thesis (1986) Utrecht University} 

\Rref{LSS}{D.A. Leites, M.V. Saveliev and V.V. Serganova, {\sl
Embeddings of Osp($N/2$) and associated non-linear supersymmetric
equations\/}, in {\sl Group Theoretical Methods in Physics\/} Vol.~1 
(Proc.~Third Yurmala Seminar USSR 22-24 May 1985),
(VNU Science Press, Utrecht, 1986) }

\Rref{FSS}{L. Frappat, A. Sciarrino and P. Sorba, Commun. Math. Phys. {\bf
121} (1989) 457}

\Rref{Dict}{L. Frappat, P. Sorba and A. Sciarrino, 
{\sl Dictionary on Lie Superalgebras}, \newline
hep-th/9607161;
preprint ENSLAPP-AL-600/96, DSF-T-30/96}

%%bosonic Toda intro
\Rref{LS}{A.N. Leznov and M.V. Saveliev, Lett. Math. Phys. {\bf 3}
(1979) 489; Commun. Math. Phys. {\bf 74} (1980) 11;
Commun. Math. Phys. {\bf 89} (1983) 59;
Acta.~Appl.~Math.~{\bf 16} (1989) 1} 

\Rref{MOP}{A.V. Mikhailov, M.A. Olshanetsky and A.M. Perelomov,
Commun. Math. Phys. {\bf 79} (1981) 473}

\Rref{PM}{P. Mansfield, Nucl.~Phys.~{\bf B208} (1982) 277; {\bf B222}
(1983) 419}

\Rref{OT}{D.I. Olive and N. Turok, Nucl.~Phys.~{\bf B257} (1986) 277;
{\bf B265} (1986) 469}

\Rref{MS}{P. Mansfield and B. Spence, Nucl. Phys. {\bf B362} (1991) 294}

\Rref{BS}{P. Bouwknegt and K. Schoutens, Phys. Rep. {\bf 223} (1993) 183}

\Rref{HR}{P. Forg\'acs, A. Wipf, J. Balog, L. Feh\'er and 
L. O'Raifeartaigh, Phys. Lett. {\bf B227} (1989) 214;
L. Feher, L. O'Raifeartaigh, P. Ruelle, I. Tsutsui and A. Wipf,
Phys. Rep. {\bf 222} (1992) 1}

\Rref{FLO}{M.D. Freeman, Phys. Lett. {\bf B261} (1991) 57\newline
A. Fring, H.C. Liao and D.I Olive, Phys. Lett. {\bf B266} (1991) 82 }

%%%%% Super Toda 
\Rref{O}{M.A. Olshanetsky, Commun. Math. Phys. {\bf88} (1983) 63} 

\Rref{LM1}{H.C. Liao and P. Mansfield, Nucl.~Phys.~{\bf B344} (1990)
696; Phys.~Lett.~{\bf B267} (1991) 188} 

\Rref{LM2}{H.C. Liao and P. Mansfield, 
Phys.~Lett.~{\bf B252} (1991) 230; {\bf B252} (1991) 237}

\Rref{EH1}{J.M. Evans and T.J. Hollowood, Nucl. Phys. {\bf B352} (1991)
723, erratum Nucl. Phys. {\bf B382} (1992) 662}
\Rref{JE}{J.M. Evans, Nucl. Phys. {\bf B390} (1993) 225 }

\Rref{EH2}{J.M. Evans and T.J. Hollowood, Phys. Lett. {\bf B293} (1992) 100}
\Rref{EM1}{J.M. Evans and J.O. Madsen, Phys. Lett. {\bf B384} (1996) 131}
\Rref{EM2}{J.M. Evans and J.O. Madsen, Phys. Lett. {\bf B389} (1996) 665}
\Rref{EM3}{J.M. Evans and J.O. Madsen, {\sl Quantum Integrability of
Coupled $N{=}1$ Super Sine/Sinh-Gordon Theories and the Lie Superalgebra
$\d(2,1;\alpha)$}, Preprint DAMTP/97-147 US-FT/35-97; hep-th/9712227}

\Rref{NM}{H. Nohara and K. Mohri, Nucl.~Phys.~{\bf B349} (1991) 253; \newline
S. Komata, K. Mohri and H. Nohara, Nucl.~Phys.~{\bf B359} (1991) 168; \newline
H. Nohara, Ann. Phys. {\bf 241} (1992) 1}

\Rref{IK}{T. Inami and H. Kanno, Commun. Math. Phys. {\bf 136} (1991) 519; 
Nucl. Phys. {\bf B359} (1991) 201} 

%%%%% Hamiltonian reduction for Superalgebras 
\Rref{SHR}{T. Inami and K-I. Izawa, Phys. Lett. {\bf B255} (1991) 521;
\newline
F. Delduc, E. Ragoucy and P. Sorba, Commun. Math. Phys. {\bf 146} (1992) 403;
\newline
L. Frappat, E. Ragoucy and P. Sorba, Commun. Math, Phys. {\bf 157} 
(1993) 499}

%%%%% B(0,n) CFT
\Rref{WI}{G.M.T. Watts, Nucl. Phys. {\bf B361} (1991) 311; \newline
K. Ito, Int. J. Mod. Phys. {\bf A7} (1992) 4885}

\Rref{Ahn}{C. Ahn, Nucl. Phys. {\bf B354} (1990) 57}
\Rref{ABL}{C. Ahn, D. Bernard and A. LeClair, Nucl. Phys. {\bf B346}
(1990) 409} 

%%%%% S-matrices for bosonic Toda
\Rref{BTS}{A.E. Arinstein, V.A. Fateev and A.B. Zamalodchikov,
Phys. Lett. {\bf B87} (1979) 389;
\newline
H.W. Braden, E. Corrigan, P.E. Dorey and Y. Sasaki,
Nucl. Phys. {\bf B338} (1990) 689;
\newline
P. Christie and G. Mussardo, Int. J. Mod. Phys. {\bf A5} (1990) 4581;
\newline
G.W. Delius, M.T. Grisaru and D. Zanon, Nucl. Phys. {\bf B382} (1992) 365}

%%%%% S-matrices for pos-def superalgebras 
\Rref{STS}{M.T. Grisaru, S. Penati and D. Zanon, 
Phys.~Lett.~{\bf B253} (1991) 357; \newline 
G.W. Delius, M.T. Grisaru, S. Penati and D. Zanon, Nucl.~Phys.~{\bf
B359} (1991) 125; Phys.~Lett.~{\bf B256} (1991) 164;
\newline 
M.T. Grisaru, S. Penati and D. Zanon, Nucl.~Phys.~{\bf B369} (1992) 373
}

\Rref{QI}{S. Penati and D. Zanon, Phys. Lett. {\bf B288} (1992) 297; 
\newline
%%%%% Quantum conserved currents in supersymmetric Toda theories IFUM-426-FT
A. Gualzetti, S. Penati and D. Zanon, 
Nucl. Phys. {\bf B398} (1993) 622} 

%%%%%% Topological Toda
\Rref{PPZ}{S. Penati, M. Pernici and D. Zanon, 
Phys.~Lett.~{\bf B309} (1993) 304;
\newline
S. Penati and D. Zanon, Nucl. Phys. {\bf B436} (1995) 265} 

%%%%% Non-unitary models
\Rref{SGpert}{T. Eguchi and S-K. Yang, Phys. Lett. {\bf B224} (1989) 373; 
{\bf B235} (1990) 282;
\newline
T.J. Hollowood and P. Mansfield, Phys. Lett. {\bf B226} (1989) 73;
\newline 
A. LeClair, Phys. Lett. {\bf B230} (1989) 103;
\newline
D. Bernard and A. LeClair, Phys. Lett. {\bf B247} (1990) 309;
\newline
F.A. Smirnov, Nucl. Phys. {\bf B337} (1990) 156;
Int. J. Mod. Phys. {\bf A4} (1989) 4213;
\newline
N. Yu. Reshetikhin and F.A. Smirnov, Commun.~Math.~Phys.~{\bf 131} (1991) 157}

\Rref{BRST}{G. Felder, Nucl. Phys. {\bf B317} (1989) 215;\newline
P. Bouwknegt, J. McCarthy and K. Pilch, Commun.~Math.~Phys.~{\bf 131}
(1990) 125} 

\Rref{CM}{J.L. Cardy and G. Mussardo, Phys.~Lett.~{\bf B225} (1989) 275}

\Rref{TH}{T.J. Hollowood, Nucl.~Phys.~{\bf B384} (1992) 523;
Int.~J.~Mod.~Phys.~{\bf A8} (1993) 947}

%%%%% N=2 SG and S-matrix
\Rref{KU}{K. Kobayashi and T. Uematsu, Phys. Lett. {\bf B264} (1991) 107;
{\bf B275} (1992) 361}

\endref
\ciao
